\documentclass[final,3p,times,twocolumn]{elsarticle}

\usepackage{amssymb}  
\usepackage{amsmath}  
\usepackage{amsthm} 
\usepackage{lineno} 

\usepackage[colorlinks,
            linkcolor=blue,
            anchorcolor=blue,
            citecolor=blue,
            urlcolor=blue]{hyperref}

\usepackage{adjustbox}
\usepackage{makecell}
\usepackage{multirow}
\usepackage{upgreek}

\graphicspath{{fig/}}

\journal{Computer Physics Communications}

\begin{document}

\begin{frontmatter}



\title{One-to-one correspondence reconstruction at the electron-positron Higgs factory}

\author[a,b]{Yuexin Wang}
\author[a,c,d]{Hao Liang}
\author[e]{Yongfeng Zhu}
\author[a,f]{Yuzhi Che}
\author[a,c]{Xin Xia}
\author[g]{Huilin Qu}
\author[e]{Chen Zhou}
\author[a,c]{Xuai Zhuang}
\author[a,c]{Manqi Ruan\corref{cor1}}
\cortext[cor1]{Corresponding author}
\ead{ruanmq@ihep.ac.cn}

\affiliation[a]{
    organization={Institute of High Energy Physics, Chinese Academy of Sciences},
    addressline={19B Yuquan Road, Shijingshan District}, 
    city={Beijing},
    postcode={100049}, 
    country={China}}
\affiliation[b]{
    organization={China Spallation Neutron Source},
    city={Dongguan},
    postcode={523803}, 
    country={China}}
\affiliation[c]{
    organization={University of Chinese Academy of Sciences},
    addressline={19A Yuquan Road, Shijingshan District},
    city={Beijing},
    postcode={100049},
    country={China}}
\affiliation[d]{
    organization={Vanderbilt University Institute of Imaging Science, Vanderbilt University Medical Center},
    city={Nashville},
    postcode={TN 37232},
    country={USA}}
\affiliation[e]{
    organization={State Key Laboratory of Nuclear Physics and Technology, School of Physics, Peking University},
    city={Beijing},
    postcode={100871},
    country={China}}
\affiliation[f]{
    organization={China Center of Advanced Science and Technology},
    city={Beijing},
    postcode={100190},
    country={China}}
\affiliation[g]{
    organization={EP Department, CERN},
    city={CH-1211 Geneva 23},
    country={Switzerland}}

\begin{abstract}
We propose one-to-one correspondence reconstruction for electron-positron Higgs factories.
For each visible particle, one-to-one correspondence aims to associate relevant detector hits with only one reconstructed particle and accurately identify its species.
To achieve this goal, we develop a novel detector concept featuring 5-dimensional calorimetry that provides spatial, energy, and time measurements for each hit, and a reconstruction framework that combines state-of-the-art particle flow and machine learning algorithms.
In the benchmark process of Higgs to di-jets, over 91\% of visible energy can be successfully mapped into well-reconstructed particles that not only maintain a one-to-one correspondence relationship but also associate with the correct combination of cluster and track, improving the invariant mass resolution of hadronically decayed Higgs bosons by 25\%. 
Performing simultaneous identification on these well-reconstructed particles, we observe efficiencies of 97\% to nearly 100\% for charged particles ($e^{\pm}$, $\mu^{\pm}$, $\pi^{\pm}$, $K^{\pm}$, $p/\bar{p}$) and photons ($\gamma$), and 75\% to 80\% for neutral hadrons ($K_L^0$, $n$, $\bar{n}$).
For physics measurements of Higgs to invisible and exotic decays, golden channels to probe new physics, one-to-one correspondence could enhance discovery power by 10\% to up to a factor of two.
This study demonstrates the necessity and feasibility of one-to-one correspondence reconstruction at electron-positron Higgs factories.
\end{abstract}

\end{frontmatter}

\section{Introduction}

In collider experiments, high-energy particles collide and generate multiple final state particles, bringing information about underlying physics laws. 
These particles interact with the detector, excite detector sensors, and generate electronic signals that are recorded into data, typically in the form of detector hits. 
Correspondingly, data processing in collider experiments consists of two main steps: interpreting detector hits into physics objects, particularly final state particles, and performing physics measurements using these physics objects.
The first step is called reconstruction, which can be regarded as the inverse process of detector hits generation.

The reconstruction establishes a mapping between visible particles and reconstructed ones, where the ultimate goal is a one-to-one (1-1) correspondence.
The visible particles include both primary particles from the interaction point (IP) and secondary particles generated through interactions with detector materials upstream of the calorimeter, as far as they create detector hits.
1-1 correspondence reconstruction aims to not only successfully reconstruct each individual particle, but also correctly identify its species (a.k.a. the particle identification, PID).
1-1 correspondence is highly beneficial for collider experiments because it provides a universal foundation for reconstructing various physics objects and can significantly enhance the reconstruction performance for jets and missing energy/momentum.

\begin{figure*}[!t]
    \centering
    \adjustbox{valign=c}{\includegraphics[scale=0.2]{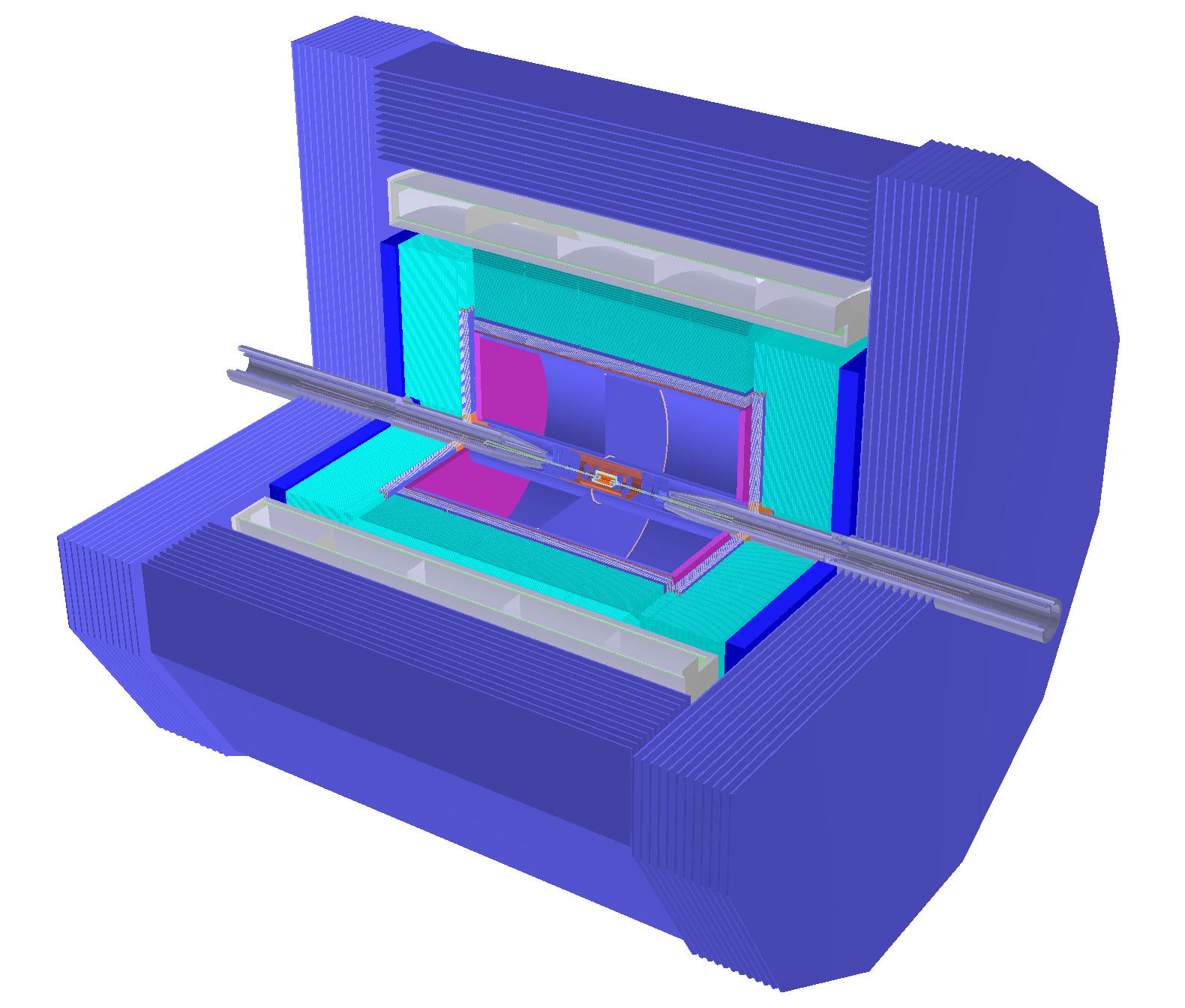}}
    \adjustbox{valign=c}{\includegraphics[scale=0.28]{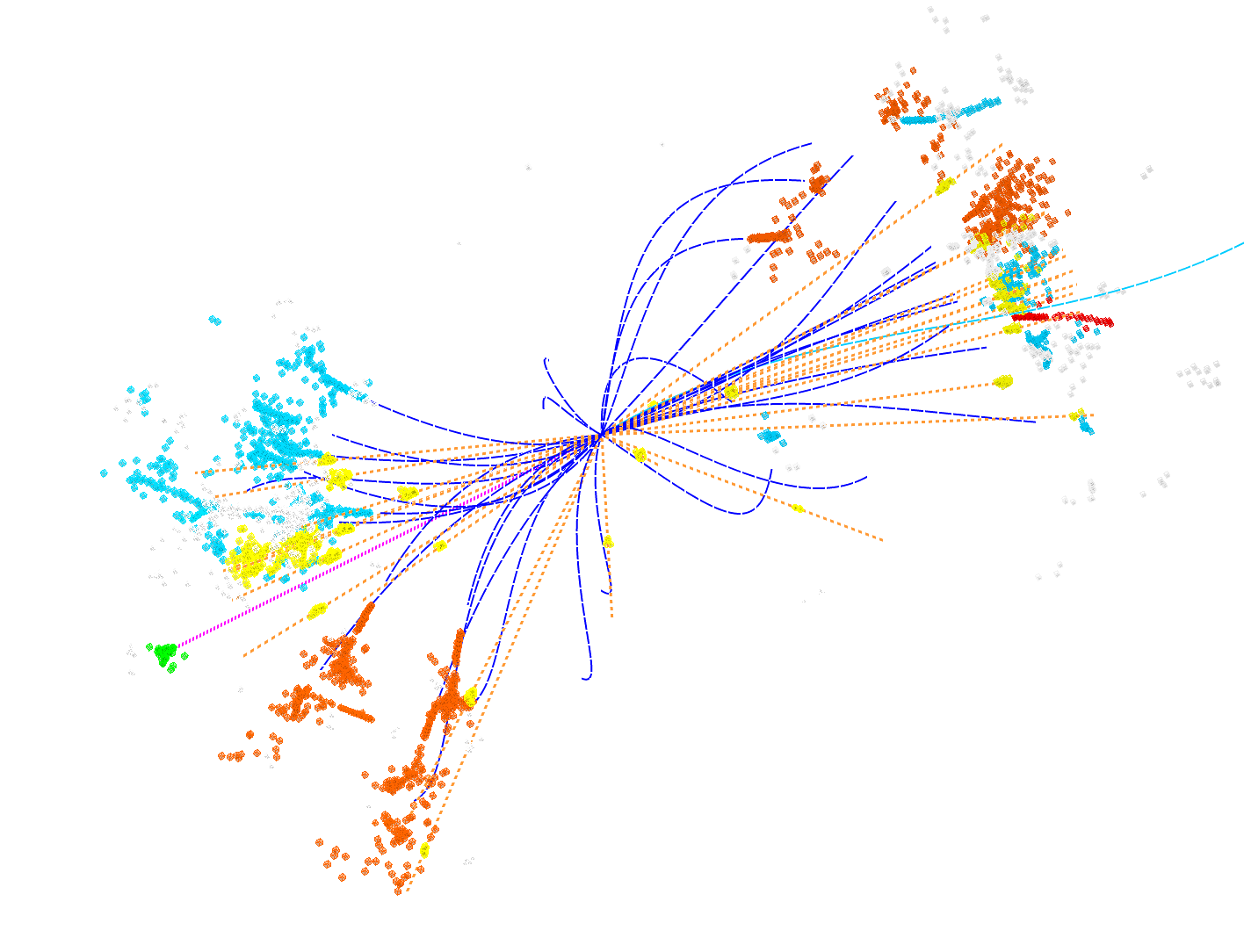}}
    \caption{AURORA detector geometry and an event display of a reconstructed $e^+e^- \to Z(\to \nu\bar{\nu})H(\to gg)$ ($\sqrt{s}$ = 240\,GeV) event.}
    \label{fig:display}
\end{figure*}

Depending on the multiplicity of visible particles and the detector configuration, achieving 1-1 correspondence could be truly challenging. 
For proton-proton and heavy ion collisions at the Large Hadron Collider (LHC)~\cite{Evans:2008zzb} or future hadron colliders, each collision can generate $\mathcal{O}(10^2\text{--}10^4)$ visible particles, making 1-1 correspondence extremely difficult, especially in the detector forward region. 
In contrast, at electron-positron colliders with center-of-mass energies ($\sqrt{s}$) around 10~GeV or lower, such as the BelleII~\cite{BelleII_Phy_2019, BelleII_TDR} and BESIII~\cite{BESIII_Phy, BESIII:2009fln} experiments, the particle multiplicity is typically below 10, and their detectors usually emphasize on the PID performance, achieving 1-1 correspondence could be much easier.
Our study focuses on the electron-positron Higgs factory~\cite{ILC_TDR_Sum, CLIC_CDR, CEPC_TDR_Acc, CEPC_CDR_Phy, FCC:2018evy}, which is regarded as the highest-priority future collider as it offers excellent opportunity to discover new physics~\cite{European_Strategy_2020, US_P5}, especially via precision measurements of the Higgs boson.
It operates at center-of-mass energies ranging from 91.2~GeV to several TeV, typically producing $\mathcal{O}(100)$ visible particles distributed in narrow jets in hadronic events.

The concept of 1-1 correspondence evolves from the Particle Flow Algorithm (PFA)~\cite{Pandora, Arbor, CMS_PFA, ALEPH_PFA,Brient:2004yq,Videau:2002sk}, with the key idea to trace each individual particle, classify and associate corresponding detector hits, and measure the particle 4-momentum using the most suitable sub-detector system. 
Originating from the ALEPH~\cite{ALEPH_PFA} experiment, the PFA has become the guiding principle for multiple conceptual detector designs at electron-positron Higgs factories~\cite{CEPC_CDR_Phy,FCC:2018evy,ILC_TDR_Det} and detector upgrade projects at HL-LHC~\cite{LHC_upgrade_CMS}.
Benefiting from the rapid development of artificial intelligence, many machine learning algorithms have been implemented to enhance PFA performance~\cite{Kakati:2024dun,Pata:2023rhh,DiBello:2022iwf}.
The current PFA performance at electron-positron Higgs factories is primarily limited by confusion~\cite{Arbor,Thomson:2009rp}, which includes the effects of fake particles, failures in track-cluster matching, particle loss due to shower overlap, etc. 
These confusion effects violate the 1-1 correspondence relationship.
Therefore, the goal of 1-1 correspondence reconstruction is to approach confusion-free particle flow reconstruction and to identify the species of all visible particles.

The performance of PFA can be quantified using the Boson Mass Resolution (BMR)~\cite{CEPC_CDR_Phy,Ruan:2018yrh}, which represents the relative mass resolution of massive bosons (such as Higgs, $Z$, and $W$ bosons) decaying into hadronic final states. 
At the electron-positron Higgs factory, a BMR smaller than 4\% is required to separate the $q\bar{q}H$ signal from $q\bar{q}Z$ background using the recoil mass method~\cite{CEPC_CDR_Phy, Yu:2020bxh}.
A smaller BMR is beneficial for all physics measurements with hadronic final states, especially for flavor physics measurements and new physics searches since they rely on accurate measurements of missing energy and momentum.
Referring to the Circular Electron-Positron Collider (CEPC) Conceptual Design Report (CDR)~\cite{CEPC_CDR_Phy}, its baseline detector design achieves a BMR of 3.7\%~\cite{Zhao:2018jiq, Ruan:2018yrh}, meeting the requirements for Higgs measurements.
In terms of PID, the CEPC baseline detector demonstrates a typical efficiency of 99.5\% and a misidentification rate around 1\% for isolated lepton identification~\cite{Yu:2017mpx, Yu:2021pxc}. 
It can also distinguish charged hadrons ($\pi^{\pm}$, $K^{\pm}$, $p/\bar{p}$) using techniques such as ${\rm d}E/{\rm d}x$ and time-of-flight (TOF) measurements~\cite{An:2018jtk, Zhu:2022hyy}.

\begin{figure*}[!t]
    \centering
    \adjustbox{valign=c}{\includegraphics[scale=0.215]{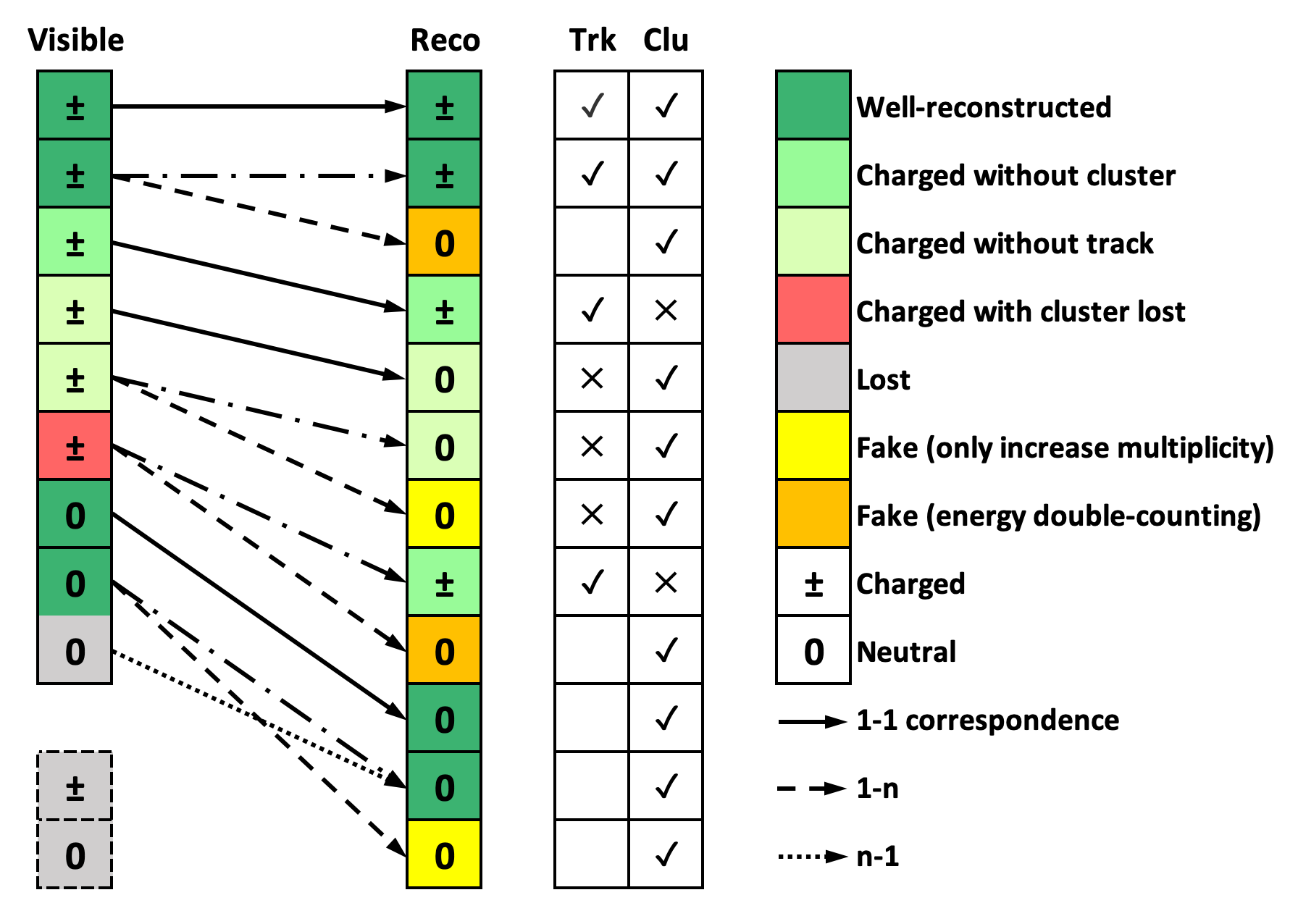}}
    \adjustbox{valign=c}{\includegraphics[scale=0.355]{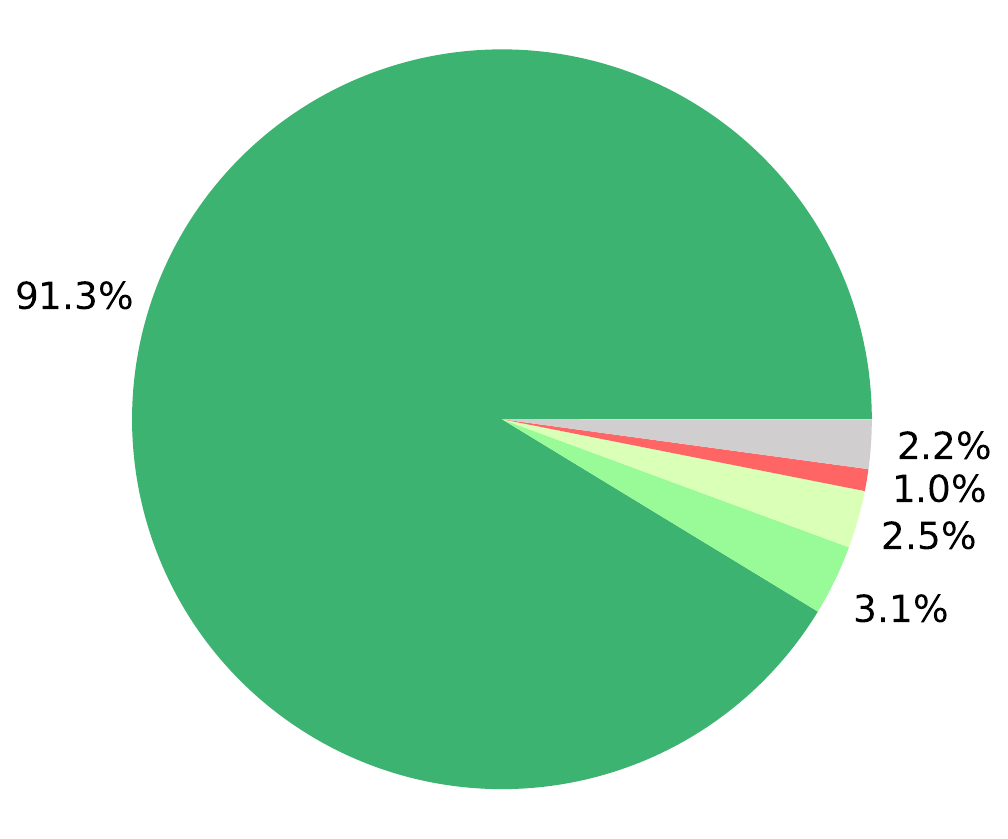}}
    \caption{
    Left: Schematic diagram of visible-reconstructed particle mapping. ``Charged without cluster/track" means a charged particle does not create enough hits to form a cluster/track (mainly due to limited detector acceptance and efficiency), while ``Charged with cluster lost" refers to cases where both a cluster and track are formed but do not match each other, resulting in fake particles.
    Right: Fraction of truth visible energy mapped into different categories.
    }
    \label{fig:mapping}
\end{figure*}

To pursue 1-1 correspondence reconstruction, we develop the AURORA (ApparatUs for RecOnstRuction with Advanced algorithm) detector concept and the PROOF (Particle Reconstruction with One-to-One correspondence at Higgs Factory) reconstruction framework, based on the CEPC CDR baseline design~\cite{CEPC_CDR_Phy}.
AURORA features a high-granularity 5-dimensional (5D) calorimetry as its core, with a unified time resolution of 100~ps for each calorimeter hit.
The left panel of Fig.~\ref{fig:display} shows a cut-away view of AURORA, with more details provided in~\ref{sec:appendix}.
Combining the CEPC baseline PFA Arbor~\cite{Arbor} and Particle Transformer (ParT)~\cite{ParT}, PROOF performs a universal identification of all reconstructed particles, simultaneously distinguishing different confusion types and identifying particle species.
ParT is a particle cloud-~\cite{Qu:2019gqs} and Transformer-based \cite{vaswani2017attention} machine learning model adapted for high-energy physics experiments, 
which has been applied in jet flavor tagging at the LHC~\cite{CMS-DP-2024-066,CMS:2024okz,Brigljevic:2024vuv} and jet origin identification (JOI) at the electron-positron Higgs factory~\cite{JOI}.
Following the convention of CEPC performance studies~\cite{CEPC_CDR_Phy,Zhao:2018jiq,GSHCAL_NIMA}, we simulate 1 million $e^+e^- \to Z(\to \nu\bar{\nu})H(\to gg)$ ($\sqrt{s}$ = 240~GeV) events (abbreviated as $ZH\to \nu\bar{\nu} gg$ in the following text) with AURORA detector and reconstruct these events with Arbor. 
A reconstructed $ZH\to \nu\bar{\nu} gg$ event is displayed in the right panel of Fig.~\ref{fig:display}, and more details on simulation and reconstruction are described in~\ref{sec:appendix}.

\section{Methods and results}

\subsection{Particle mapping}

\begin{figure*}[!t]
    \centering
    \adjustbox{valign=c}{\includegraphics[scale=0.31]{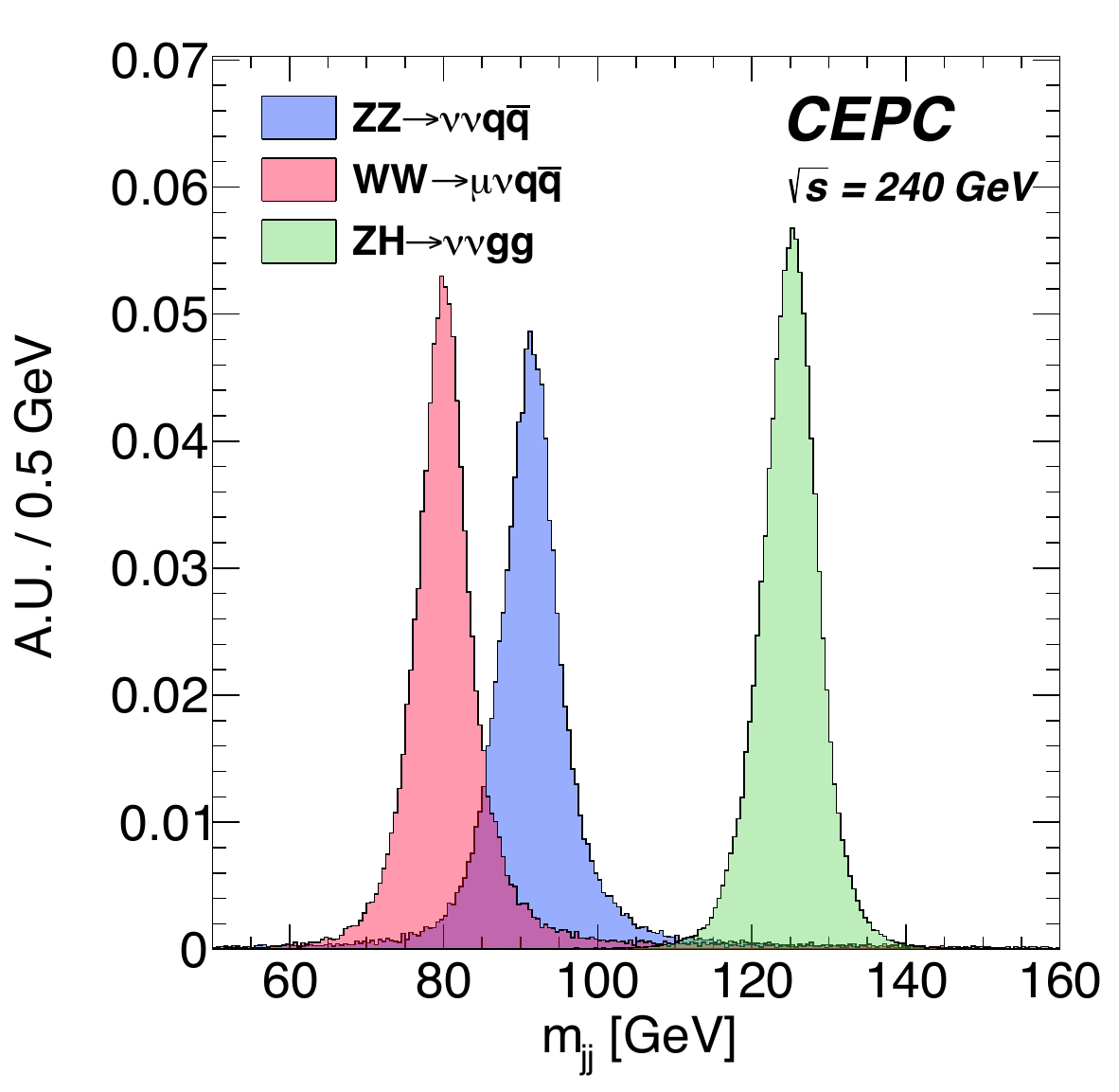}}
    \adjustbox{valign=c}{\includegraphics[scale=0.255]{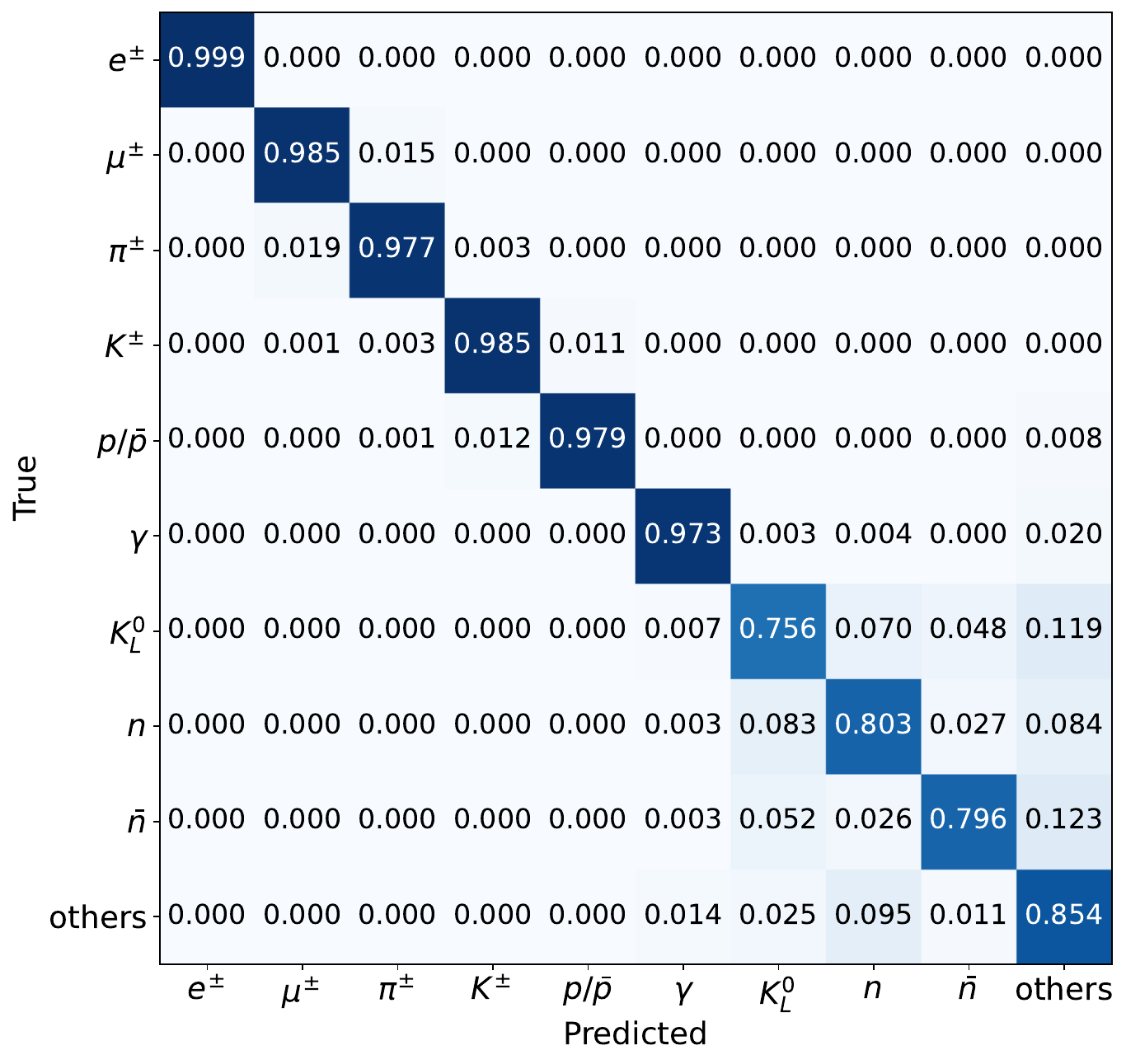}}
    \caption{
    Left: Invariant mass distributions of hadronically decayed Higgs, $W$, and $Z$ bosons derived by the combination of AURORA and PROOF to approach 1-1 correspondence reconstruction.
    Right: Confusion matrix of \textbf{well-reconstructed} particles identification.
    Both are after removing identified fake particles.
    }
    \label{fig:matrix}
\end{figure*}

Using truth-level information in the Monte Carlo (MC) simulation, we establish a mapping between simulated visible particles and reconstructed ones, as illustrated in the left panel of Fig.~\ref{fig:mapping}. 
The mapping consists of three main categories:
\begin{itemize}
    \item 1-1 correspondence (solid arrows in Fig.~\ref{fig:mapping}): 
    one visible particle is mapped to only one reconstructed particle.
    This category consists of three types, represented in shades of green in Fig.~\ref{fig:mapping}.
    One type is \textbf{well-reconstructed}, where a cluster is a must for neutral particles, and both track and cluster are required for charged ones.
    The other two types correspond to charged particles that generate only track or cluster, mainly caused by the limitations of detector acceptance (e.g. in the forward region) and efficiency (e.g. dead zone or energy/momentum measurement thresholds). 
    
    \item Fake particles (in orange and yellow in Fig.~\ref{fig:mapping}): occur when one visible particle is mapped to multiple reconstructed particles.
    Due to detector inefficiency and particle interactions with upstream materials, the calorimeter hits generated by a single particle may be grouped into multiple clusters. 
    For a charged particle, its track could be matched to only part of or even none of these clusters, while the remaining clusters are reconstructed as nearby extra neutral particles. 
    These neutral particles that originate from charged particle shower fragments and un-associated clusters, represented in orange in Fig.~\ref{fig:mapping}, can cause double-counting in the total reconstructed energy.
    For simplicity, these neutral particles are called fake particles in this study. 
    
    \item Lost particles (in gray in Fig.~\ref{fig:mapping}): visible particles with no corresponding reconstructed particles.
    Particles can be lost due to limited detector acceptance\footnote{In this article, we define visible particles as those that create detector hits. On the other hand, the limited detector acceptance causes particle loss, i.e., AURORA has an overall geometry acceptance of $|\cos\theta| \approx 0.99$, where it cannot record particles traveling in the forward direction. Since particle loss due to detector acceptance certainly impacts the overall physics measurement, we list these particles in dashed boxes in Fig.~\ref{fig:mapping} to complete the picture and also consider them when calculating visible energy fractions in the following text.}, selection criteria, and shower merging. 
    Selection criteria are applied to improve the reconstruction performance, such as vetoing fake particles, but inevitably they could exclude genuine particles, usually low-energy ones.
    Shower merging (dotted arrow in Fig.~\ref{fig:mapping}) is another major cause of particle loss, particularly within high-energy jets where nearby or overlapping calorimeter showers can be merged into one cluster, especially at limited calorimeter granularity. 
    For instance, the reconstruction of $\pi^0\to\gamma\gamma$ can be particularly sensitive to shower merging.
\end{itemize}
According to this mapping with truth information, we label reconstructed particles into 15 types, including 10 types of visible particles ($e^{\pm}$, $\mu^{\pm}$, $\pi^{\pm}$, $K^{\pm}$, $p/\bar{p}$, $\gamma$, $K_L^0$, $n$, $\bar{n}$, and others including $K_S^0$ and $\Lambda$ that do not yet decay within the tracker volume) and 5 types of confusion (such as different kinds of fake particles, charged particle without track, etc).
Each reconstructed particle is characterized by 55 observable variables as inputs to ParT.
More details can be found in~\ref{sec:appendix}.

\subsection{Fake particles and BMR}

The quantitative analysis~\cite{Manqi_CEPCWS_Chicago_2019} shows that fake particles are the leading contribution to the BMR.
In our simulation, on average, a $ZH\to \nu\bar{\nu} gg$ event has 80 visible particles with a total energy of 135~GeV at the truth level, while the reconstructed fake particles lead to a double-counted energy of 6~GeV, severely degrading the BMR.
PROOF can identify the fake particles with a typical efficiency of 77\% and a purity of 97.5\%, 
consequently suppressing the average double-counted energy from fake particles to 0.7~GeV (by nearly one order of magnitude) at the minor cost of mis-excluding visible particles with a total energy on the order of 0.5~GeV.
Consequently, PROOF achieves a BMR of 2.75\% (see the left panel of Fig.~\ref{fig:matrix}), where the effects of fake and mis-excluded particles get balanced.
Compared to the CEPC CDR~\cite{Lai:2021rko}, the BMR is improved by 25\%, where 10\% is from the detector geometry optimization to AURORA~\cite{Hu:2023dbm}, and 15\% is from the fake particle suppression.
The contributions from the three categories of visible-reconstructed particle mapping to the BMR are not proportional to their corresponding energy fractions.
The contribution from the 1-1 correspondence category primarily stems from the detector resolution, which is orders of magnitude lower compared to the total energy of particles within this category.
Meanwhile, the contributions from the fake and lost categories are comparable to their corresponding total energies.
We found that the 1-1 correspondence category contributes to half of the BMR, while the fake and lost categories each contribute a quarter, combined quadratically.

\begin{figure*}[!t]
    \centering
    \includegraphics[width=0.39\textwidth]{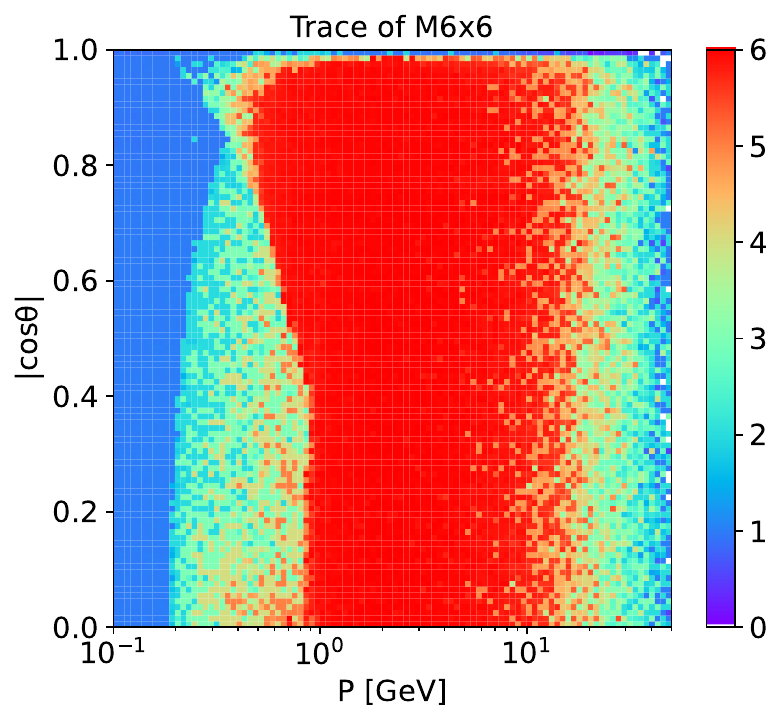}
    \includegraphics[width=0.39\textwidth]{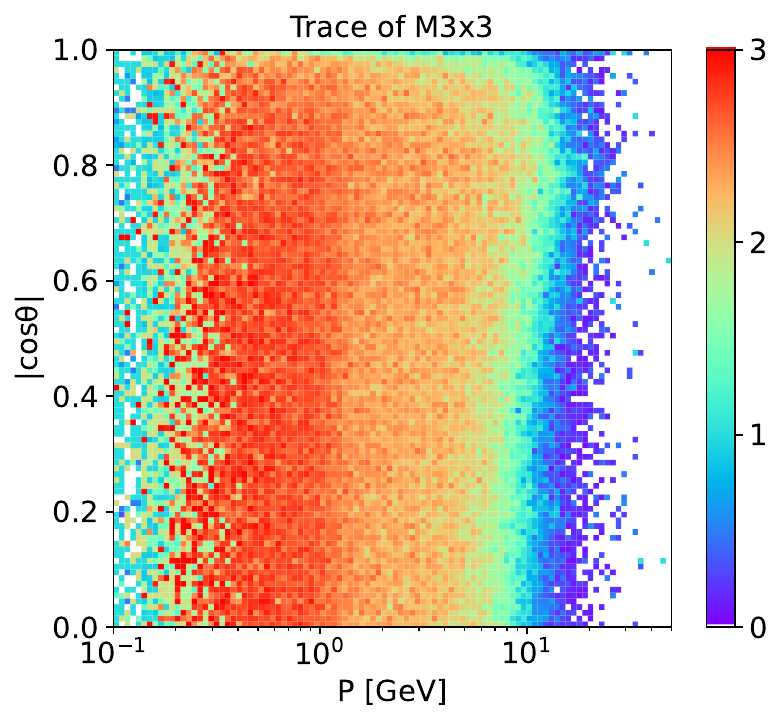}
    \caption{Traces of $6 \times 6$ ($e^{\pm}$, $\mu^{\pm}$, $\pi^{\pm}$, $K^{\pm}$, $p/\bar{p}$, $\gamma$) (left) and $3 \times 3$ ($K_L^0$, $n$, $\bar{n}$) (right) sub-matrices as function of momentum and polar angle $|\cos\theta|$.}
    \label{fig:trace_M6x6}
\end{figure*}

Referring to the truth level energy of visible particles, we calculate the fraction of total visible energy corresponding to different categories of the mapping.
Ignoring the remaining fake particles\footnote{The remaining fake particles could violate the 1-1 correspondence relationship, as one visible particle could be mapped to one leading reconstructed particle together with one or even several fake particles. 
However, as the energy of the fake particle is usually much lower than the leading one, the fake particle only has a tiny impact on the reconstruction and identification of the original visible particle. 
Meanwhile, the total energy of remaining fake particles only contributes to a sub-percentage level of total visible energy.
Therefore, ignoring these remaining fake particles in our study.}, we observe that over 91\% of the total visible energy is mapped to \textbf{well-reconstructed} particles, approximately 4\% to charged particles without clusters (with over three-fourths of this due to calorimeter acceptance or inefficiency), 2.5\% to charged particles without tracks, and 2.2\% of the energy is lost due to limited detector acceptance, shower merging, and mis-excluded genuine particles, as illustrated in the pie chart in Fig.~\ref{fig:mapping}.
The detector acceptance dominates the confusion of charged particles without clusters or tracks. 
Summing the energy fraction of \textbf{well-reconstructed} particles and confusion type purely originating from detector acceptance, we consider that nearly 95\% of the total visible energy maintains the 1-1 correspondence relationship.

\subsection{Particle identification}

Focusing on the well-reconstructed particles of 1-1 correspondence type, the PID performance of PROOF is shown in the confusion matrix in the right panel of Fig.~\ref{fig:matrix}.
The identification efficiencies (diagonal matrix elements) range from 97\% to nearly 100\% for charged particles and photons, and from 75\% to 80\% for three types of neutral hadrons.
Since the PID performance depends on particle kinematics,
we extract the traces (i.e. sums of diagonal elements) of sub-matrices in different particle momentum and polar angle ranges, as shown in Fig.~\ref{fig:trace_M6x6}. 
For photons and charged particles, a high-trace plateau is observed within the momentum range of 0.8 to 20~GeV and $|\cos\theta| < 0.96$.
Inefficiencies primarily arise due to limited cluster-level information in the low-momentum region and degraded PID performance in high-momentum and detector-forward regions. 
Similar behavior is observed for neutral hadrons, which shows that the trace is approaching 3 in a limited phase space, indicating that even neutral hadrons can be efficiently separated.
These observations lead to the conclusion that 1-1 correspondence is feasible at future electron-positron Higgs factories.

It should be noted that many simplifications have been made to facilitate this analysis.
To fully realize 1-1 correspondence reconstruction and maximize its impact on physics exploration, dedicated studies are required in the future. 
Secondary particles can significantly affect reconstruction and physics measurements.
For example, electron-positron pairs from photon conversions and muons from hadron decays could contribute to the background in measurements that rely on (semi-)leptonic
decays. Secondary nucleons, generated in the interactions between primary particles and upstream materials, as well as backscattering, could impact the BMR.
The reconstruction of the primary vertex is crucial for TOF measurements and affects PID performance.
The impact of beam-induced background needs to be analyzed and presumably ameliorated. 
Event building at high event rates also needs to be developed. 
Detector optimization should be systematically performed, including the quantification of performance specification (particularly the TOF resolution of the calorimeter and ${\rm d}E/{\rm d}x$ performance of the gaseous tracker), the survey of available and emerging technologies, and the geometry optimization. 
Modeling and verifying detailed detector responses are crucial for controlling systematic uncertainties, which is vital for precision measurements.

\section{Prospects}\label{sec:impact}

This section discusses the prospects for high-level reconstruction and physics measurements with an ideal 1-1 correspondence reconstruction.

\begin{figure*}[!htbp]
    \centering
    \adjustbox{valign=t}{\includegraphics[scale=0.3]{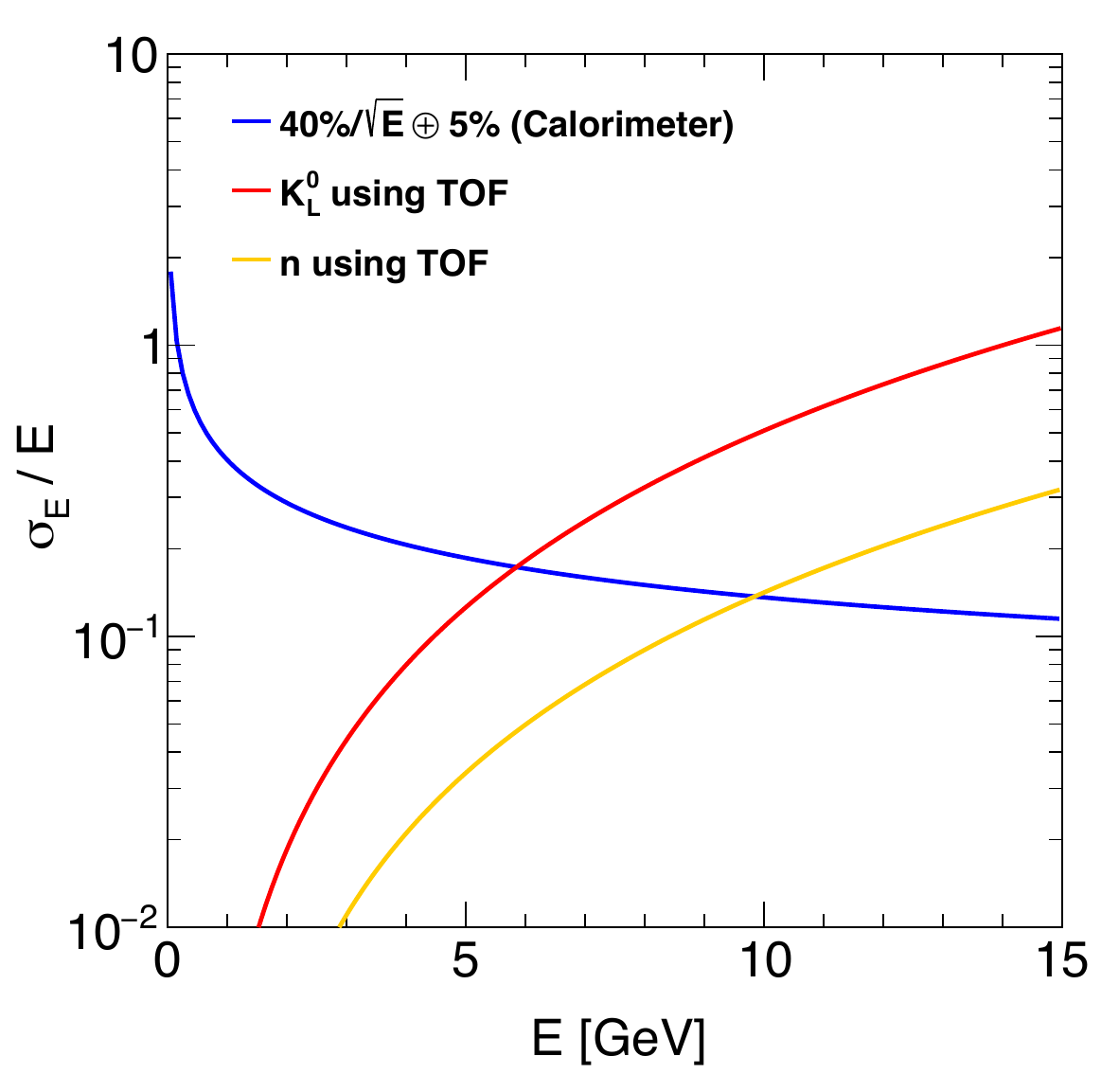}}
    \adjustbox{valign=t}{\includegraphics[scale=0.43]{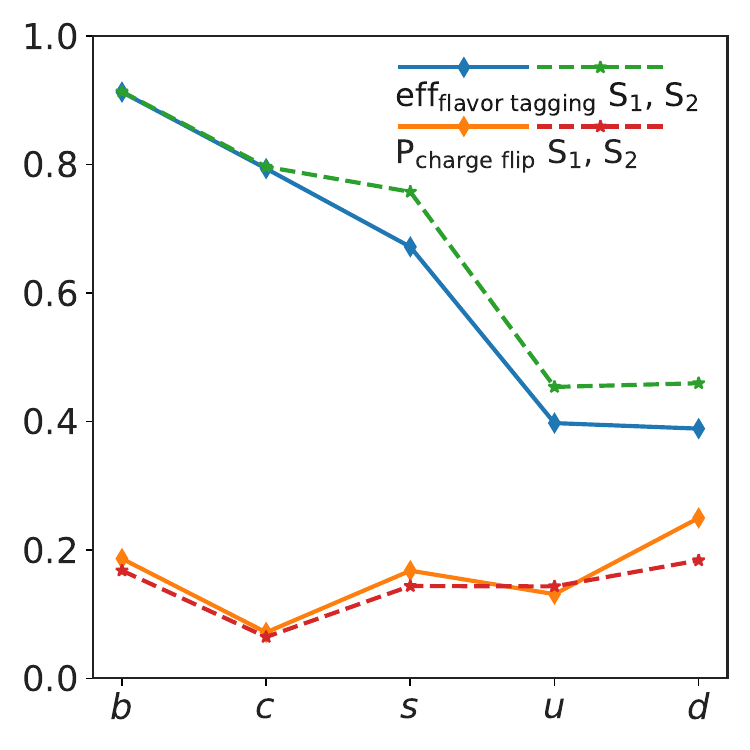}}
    \\
    \includegraphics[scale=0.7]{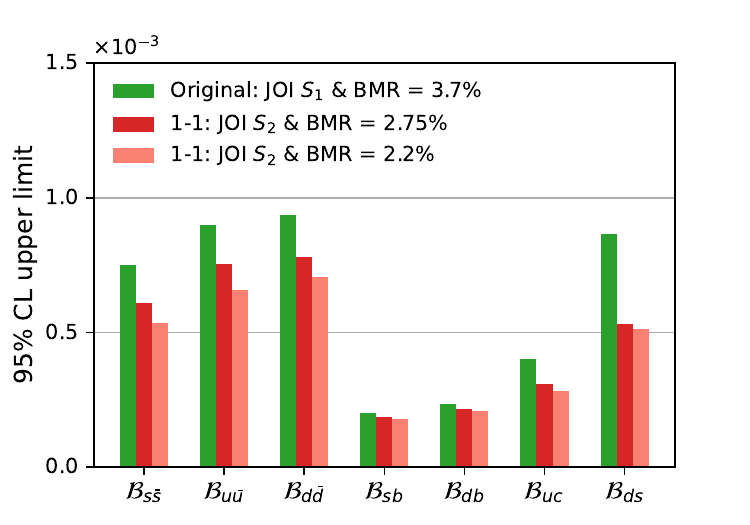}
    \caption{
    Top left: Neutral hadron energy resolution using calorimeter with direct energy sum of each hit~\cite{Hu:2023dbm,GSHCAL_talk} (blue line, derived by the full detector simulation) or TOF measurement with time resolution assumed to be 10 ps at particle level (red and yellow lines).
    Top right: Jet origin identification performance (characterized by flavor tagging efficiencies and charge flip rates) in two scenarios: Scenario 1 (S$_1$) assumes perfect identification of leptons and charged hadrons, while Scenario 2 (S$_2$) extends S$_1$ by further assuming perfect identification of $K^{0}_{L}/K^{0}_{S}$, $n/\bar{n}$, and $\Lambda/\bar{\Lambda}$.
    Bottom: Anticipated upper limits on branching ratios of Higgs exotic and rare decays with different performances of JOI and BMR.
    }
    \label{fig:JOI}
\end{figure*}

Since 1-1 correspondence reconstruction provides the particle mass information via PID, the energy of visible particles can be determined by the TOF.
5D calorimetry provides TOF measurements for both charged and neutral particles.
For charged particles, their energies/momenta can be determined by three methods: shower energy via calorimeter, track momentum via tracker, and TOF via calorimeter, while the TOF method has a comparative advantage in the forward region. 
For neutral particles, their energies can be determined by both the shower energy and the TOF via calorimeter, with TOF showing a significant advantage for low-energy neutral hadrons, as demonstrated in the top left panel of Fig.~\ref{fig:JOI}.
Therefore, by combining these measurements of particle energy, such as selecting the most suitable method according to the particle species and their kinematics, 
we can significantly improve the energy/momentum resolution of particles, and consequently improve the BMR.
Considering the improvement in particle energy measurement can also benefit pattern recognition and confusion control, we estimate that this approach could further enhance the BMR by 15--20\%, meaning that a BMR of 2.2--2.4\% could, in principle, be achieved.

1-1 correspondence reconstruction can also significantly enhance the JOI performance.
JOI is the procedure to determine the type of quark/gluon (11 types are typically considered: $b$, $\bar{b}$, $c$, $\bar{c}$, $s$, $\bar{s}$, $u$, $\bar{u}$, $d$, $\bar{d}$, $g$) from which a jet originates.
A recent study realizes JOI with the CEPC CDR baseline detector~\cite{JOI}.
It can simultaneously identify $b$, $c$, $s$ quarks with efficiencies of 70--90\% and $u$, $d$ quarks with efficiencies of 40\%, 
meanwhile, the charge flip rates of quarks and anti-quarks are controlled to be 10--20\%, with an ideal lepton and charged hadron identification.
JOI significantly boosts the accuracy of flavor-sensitive physics measurements, such as $H\to s\bar{s}$ and $H\to sb$ decays~\cite{JOI}, CKM matrix element $|V_{cb}|$~\cite{Liang:2024hox}, $CP$-violating phase $\phi_s$~\cite{Li:2022tlo}, and weak mixing angle~\cite{Zhao:2022lyl} measurements. 
1-1 correspondence provides the identification of $K^{0}_{L}$, $n$, and $\bar{n}$, while in principle the $K^{0}_{S}$ and $\Lambda$ could also be identified~\cite{Zheng:2020qyh}, therefore, providing the PID information of all these neutral hadrons, we observe a significant improvement on JOI performance especially the identification performance of $u/d/s$ quarks, as shown in the top right panel of Fig.~\ref{fig:JOI}.

BMR and JOI are key performances of event reconstruction at electron-positron Higgs factories. 
Their improvements will benefit most physics measurements with hadronic final states. 
For example, the Higgs to invisible final states is a key portal to detect dark matter at Higgs factories. 
The original upper limit at 95\% confidence level is quantified to be $\mathcal{O}(0.1\%)$~\cite{CEPC_Snowmass} with a BMR of 3.7\%.
Improving the BMR to 2.75\%, or even 2.2\%, will consequently enhance the upper limit by $\mathcal{O}(10\%)$~\cite{Manqi_HKIAS_2024}.
Higgs exotic and rare hadronic decays are also sensitive probes to new physics. 
Using JOI technology, these decay modes can be typically limited to $10^{-4}$ to $10^{-3}$ using $\nu\bar{\nu}H$ and $\ell\ell H$ processes~\cite{JOI}.
With the improvement in JOI alone, the constraints could be enhanced by 10\% to up to a factor of two.
Considering also the improved BMR, the sensitivities are expected to be further enhanced by $\mathcal{O}(10\%)$, see the bottom panel of Fig.~\ref{fig:JOI}.

1-1 correspondence can be regarded as a high-efficient information compression that abstracts the entire physics event into $\mathcal{O}(100)$ reconstructed particles, each characterized by approximately $\mathcal{O}(10)$ observables.
It enables new analysis methodology to distinguish the signal from background events using all these observables directly. 
Compared to conventional methodologies, for example the mainstream cut-based and BDT methods that rely on $\mathcal{O}(10)$ of human-defined variables, this new method describes physics events in a much larger parameter space (with the dimension increased by roughly 2 orders of magnitude), thereby significantly enhancing the signal/background separation. 
A toy analysis demonstrates the effectiveness of this new method in distinguishing the $ZH$ signal from $ZZ$ and $WW$ backgrounds in fully hadronic final states, achieving a performance improvement by more than a factor of 2~\cite{Manqi_CSI_talk}.
This method can also be applied to detector monitoring and calibration, providing much more information to control the experiment systematic uncertainties. 
It should be noted that 1-1 correspondence-based analysis relies on accurate MC simulations to provide large statistics of labeled samples. Therefore, the development and validation of suitable MC tools have become critical and should be iteratively refined alongside data analysis, theoretical calculations, and modeling.

\section{Summary}\label{sec:summary}

Regarding the entire detector signal recording and event reconstruction as the process to establish a mapping between visible and reconstructed particles, 1-1 correspondence represents the ultimate goal of an isomorphic mapping. 
It characterizes events with observables associated with individually reconstructed particles, enabling direct and holistic signal selection algorithms that can significantly enhance the discovery power.
This study demonstrates the feasibility of 1-1 correspondence reconstruction at the electron-positron Higgs factory.
By employing innovative detector designs, such as high-granularity 5D calorimetry, alongside advanced algorithms like PFA and machine learning, it is possible to efficiently suppress and identify fake particles---the leading source of confusion that violates the 1-1 correspondence relationship, leading to a 25\% improvement in the invariant mass resolution of hadronically decayed Higgs bosons (i.e. BMR).
Using the same machine learning model, nine types of particles ($e^{\pm}$, $\mu^{\pm}$, $\pi^{\pm}$, $K^{\pm}$, $p/\bar{p}$, $\gamma$, $K_L^0$, $n$, $\bar{n}$) can be simultaneously distinguished with high efficiencies, greatly expanding and unifying traditional PID methods.
1-1 correspondence reconstruction also shifts the bottlenecks in experiment design, event reconstruction, and physics measurements.  
This study shows that the primary bottleneck in particle flow reconstruction, namely confusion, can be reduced by one order of magnitude, while the current limitations shift to detector acceptance and neutral particle identification. 
This study also urges the necessity of high-precision MC tools, which require profound understanding of hadronization processes, particle-matter interactions, detector responses, etc.
The study also indicates that 1-1 correspondence represents an emerging paradigm for event reconstruction at the high-energy frontier. 
Achieving this paradigm relies on state-of-the-art detector design, innovative reconstruction algorithms, and artificial intelligence, potentially leading to a significantly deeper understanding of physics events and the underlying physics principles.

\section*{CRediT authorship contribution statement}

\textbf{Yuexin Wang:} Investigation, Methodology, Data curation, Formal analysis, Visualization, Writing - original draft, Writing – review and editing.
\textbf{Hao Liang:} Investigation, Methodology, Formal analysis, Visualization, Writing – review and editing.
\textbf{Yongfeng Zhu:} Investigation, Methodology, Formal analysis, Visualization, Writing – review and editing.
\textbf{Yuzhi Che:} Investigation, Methodology, Writing – review and editing.
\textbf{Xin Xia:} Data curation, Validation, Writing – review and editing.
\textbf{Huilin Qu:} Software, Writing – review and editing.
\textbf{Chen Zhou:} Writing – review and editing.
\textbf{Xuai Zhuang:} Writing – review and editing.
\textbf{Manqi Ruan:} Conceptualization, Methodology, Writing - original draft, Writing – review and editing, Supervision, Funding acquisition.

\section*{Declaration of competing interest}

The authors declare that they have no known competing financial interests or personal relationships that could have appeared to influence the work reported in this paper.

\section*{Acknowledgements}

We would like to express our sincere gratitude to Jianchun Wang for his continuous support throughout this work. In addition, we thank Hengyu Wang for his study on HCAL resolution and thank Jianfeng Jiang for his study on event generator, which provided important references for our research. We are also grateful to Yuxuan Zhang for his review of the manuscript and valuable suggestions.
This work is supported in part by the National Natural Science Foundation of China under grant No. 12042507, the National Key R\&D Program of China under Contracts No. 2022YFE0116900, and the National Key Program for S\&T Research and Development (2023YFA1606300). Additional support is provided by the Fundamental Research Funds for the Central Universities, Peking University.

\section*{Data availability}

The code used to produce the samples and perform the analysis is available at \url{https://code.ihep.ac.cn/wangyuexin/1-1-correspondence-reconstruction.git}


\appendix

\section{Appendix}
\label{sec:appendix}

\subsection{AURORA detector}

\begin{figure}[!t]
    \centering
    \includegraphics[width=.38\textwidth]{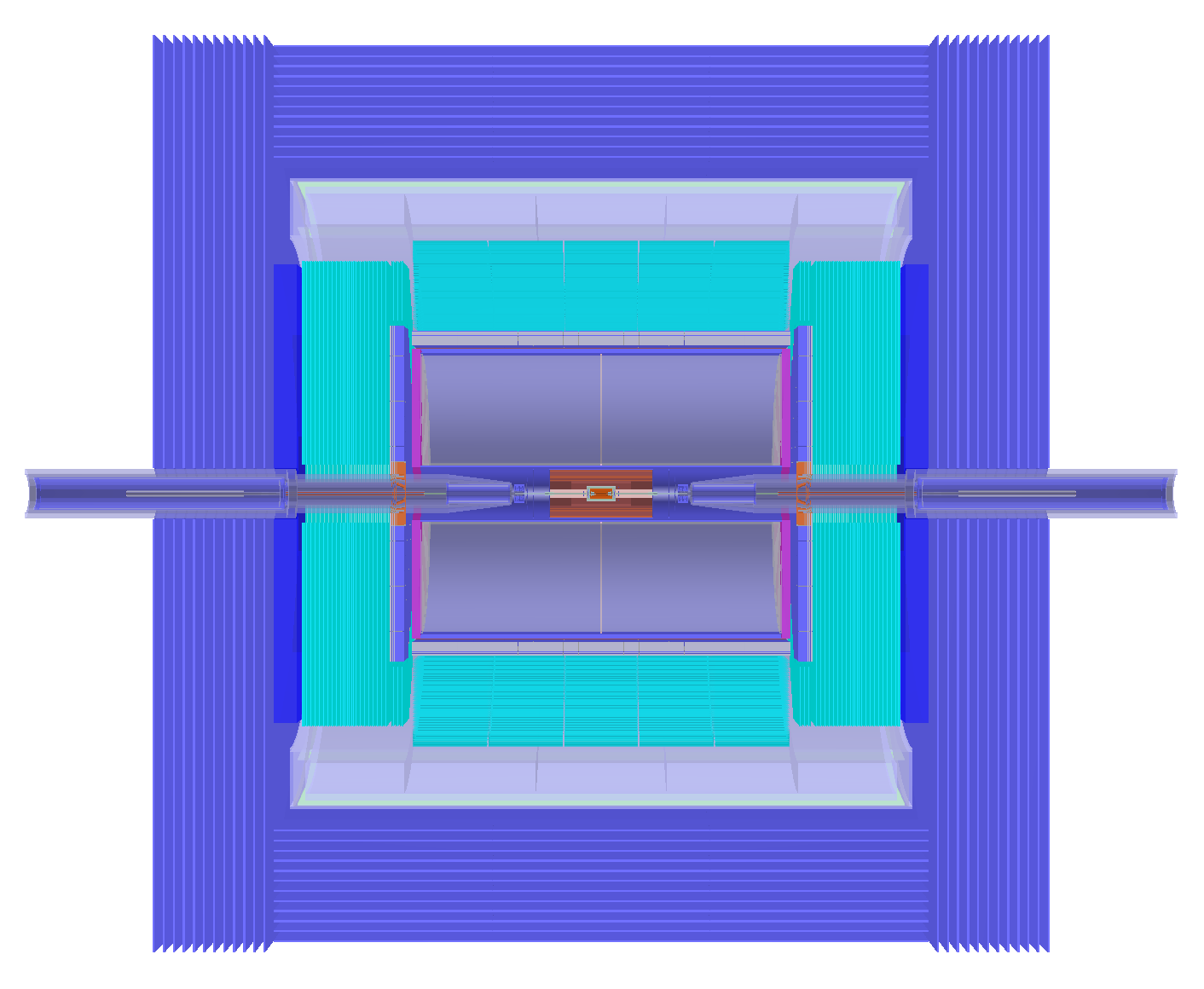}
    \includegraphics[width=.38\textwidth]{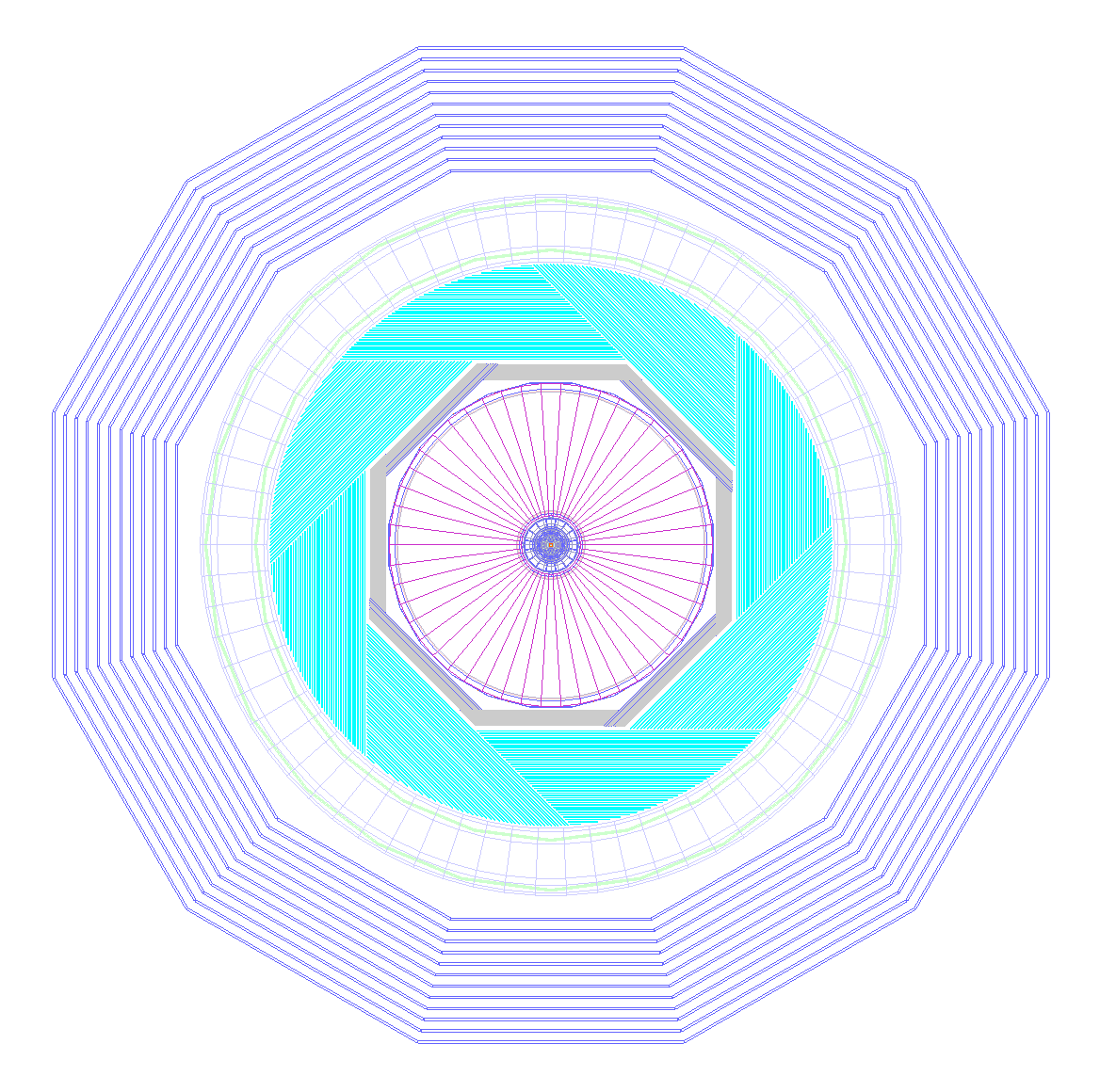}
    \caption{The $r$-$z$ (left) and $r$-$\phi$ (right) views of the AURORA detector geometry.}
    \label{fig:geo}
\end{figure}

AURORA, the conceptual detector used in this study, is a PFA-oriented detector design developed from the CEPC CDR baseline~\cite{CEPC_CDR_Phy,Liang:2022qxn}.
It has an outer radius of 5.2~m, a length of 10.5~m, and a total weight of approximately 5000~tons, with over 65\% of the weight contributed by the Yoke and nearly 25\% by the HCAL.
Fig.~\ref{fig:geo} displays two side views of AURORA, and specific geometry parameters are summarized in Table~\ref{tab:geo_parameter}.
In general, AURORA features a high-granularity calorimetry with both ECAL and HCAL, a high-precision tracker system with a low material budget, a high-precision vertex detector, and a large solenoid that encloses ECAL and HCAL.
Compared to the CDR baseline, the major change is the HCAL, where we replace the Glass Resistive Plate Chamber (GRPC) digital HCAL~\cite{CEPC_CDR_Phy} with the Glass-Scintillator HCAL~\cite{GSHCAL_NIMA} and increase the HCAL thickness from 5 to 6 nuclear interaction lengths ($\lambda$) to reduce the longitudinal leakage.
In addition, for both ECAL and HCAL, we assume each calorimeter cell can provide a time resolution of 100~ps, which is used for the PID and PFA confusion identification.

\begin{table*}[!t]
    \centering
    \caption{AURORA detector geometry parameters.}    
    \label{tab:geo_parameter}
  	\resizebox{1.\textwidth}{!}{
        \begin{tabular}{ccccccccc}
        \hline
        Sub-detector & \makecell{Thickness\\ (mm)}
        & \makecell{Inner radius\\ (mm)} & \makecell{Outer radius\\ (mm)} & \makecell{Length\\ (mm)} & \makecell{Volume\\ (m$^3$)} & \makecell{Transverse\\ cell size} & \#Layers & \#Channels \\
        \hline
        
        Vertex & -
        & - & 16--60 & 125--250 & - & $25 \times 25$ $\upmu$m$^2$ & 6 & $5.3\times10^8$ \\
        \hline
        
        \makecell{Si-strip\\ Tracker} & -
        & - & \makecell{155\\ 300\\ 1810} & \makecell{736\\ 1288\\ 4600} & - & 20 $\upmu$m $\times$ 2 cm & 3 & $3.0\times10^{7}$ \\
        \hline
        
        TPC & -
        & 300 & 1800 & 4700 & 47 & $1 \times 6$ mm$^2$ & 220 & $2.9\times10^6$ \\
        \hline
        
        ECAL & 173
        & 1845 & 2018 & 5250 & 15 & $1 \times 1$ cm$^2$ & 30 & $2.5 \times 10^7$ \\
        \hline
        
        HCAL & 1145
        & 2072 & 3250 & 7590 & 180 & $2 \times 2$ cm$^2$ & 48 & $1.8 \times 10^7$ \\
        \hline
        
        Solenoid & 700
        & 3275 & 3975 & 7750 & 120 & - & - & - \\
        \hline
        
        Yoke & 1200
        & 4000 & 5200 & 10500 & 470 & - & - & - \\
        
        \hline
        \end{tabular}
    }

\end{table*}

\subsection{Monte Carlo simulation and reconstruction}

A total of 1 million $ZH\to \nu\bar{\nu} gg$ events are generated and simulated using the Monte Carlo method.
The simulation is conducted within the CEPC software framework~\cite{CEPC_CDR_Phy}, utilizing Whizard1.95~\cite{Whizard} and Pythia6.4~\cite{Pythia6} for event generation, MokkaPlus~\cite{MoradeFreitas:2004sq,Mokka_CEPCNote} for detector simulation, and Arbor~\cite{Arbor,Ruan:2018yrh} for particle flow reconstruction.
Using tracks and calorimeter hits as inputs, Arbor groups calorimeter hits into clusters based on the shower’s tree-like topology. With the high granularity of the AURORA calorimeter, Arbor can efficiently separate nearby showers.

For future studies, we also plan to perform comparative analysis with different generators, such as Pythia8~\cite{Bierlich:2022pfr}, MadGraph~\cite{Alwall:2014hca}, and Herwig~\cite{Bellm:2015jjp}, to quantify the impact of different hadronization models on advanced reconstruction algorithms, which is crucial for the relevant uncertainty control.

\subsection{Particle mapping}

The visible-reconstructed particle mapping is constructed using the truth links recorded in the simulation sample. 
These truth links connect each tracker/calorimeter hit with the particles that excite the hit. 
The hits are also associated with reconstructed particles through the reconstruction procedure.
The mapping is thus established via hits.
By analyzing the mapping, all reconstructed particles are labeled into 15 types:
\begin{itemize}
    \item 10 types of 1-1 correspondence (in shades of green in Fig.~\ref{fig:mapping}): $e^{\pm}$, $\mu^{\pm}$, $\pi^{\pm}$, $K^{\pm}$, $p/\bar{p}$, $\gamma$, $K_L^0$, $n$, $\bar{n}$, and others such as $K_S^0$ and $\Lambda$.
    
    \item Charged particles with no track (in light green in Fig.~\ref{fig:mapping}): Since a track is an indispensable feature that provides more precise measurement and key information for identifying charged particles, the absence of a track will undoubtedly affect the identification of charged particles. 
    We therefore classify this case as a separate category.
    
    \item Fake particles from charged shower fragments (in orange in Fig.~\ref{fig:mapping}).
    
    \item Fake particles from neutral shower fragments (in yellow in Fig.~\ref{fig:mapping}).
    
    \item Multi-track: more than one charged reconstructed particle (track) is mapped to one visible particle. Considering that the proportion of this type is less than 0.03\%, we omit this case in the mapping schematic diagram for simplicity.
    
    \item In Arbor reconstruction, a small fraction ($<0.1\%$) of reconstructed particles cannot be mapped to any visible particle. These particles are virtually reconstructed through Arbor's supplementary ``energy flow" strategy to balance energy during track and cluster matching. Since this is an algorithm-dependent case, it is also omitted in the mapping schematic diagram.
\end{itemize}

\subsection{ParT}

ParT~\cite{ParT} is used to classify reconstructed particles through an encoder-only Transformer architecture. This model processes both individual particle features and pairwise particle interaction features. The particle features are first embedded with a 3-layer multi-layer perceptron (MLP) and then passed through 10 layers of multi-head self-attention, where the pairwise interaction features introduce a pre-softmax bias to the attention mechanism. The final particle embeddings, obtained from the last attention layer, are further processed by three fully connected layers, each with 128 neurons and a dropout rate of 0.1. A final linear layer projects the embeddings into a space with dimensions corresponding to the number of classes, generating a set of scores that represent the classification probabilities for each particle.

\subsubsection{Training setup}

Each reconstructed particle is characterized by 55 observable variables at the particle, cluster, and track levels, as listed in Table~\ref{tab:input_var}. 
These variables serve as input features for ParT, which outputs the likelihoods for different types of reconstructed particles. 
The complete sample of 1 million reconstructed $ZH\to \nu\bar{\nu} gg$ events are split into three independent sets in a ratio of 6:2:2 for training, validation, and evaluation of the ParT model.
A ParT model with approximately 2.2 million parameters is constructed and trained for 30 epochs.

\begin{table*}[!t]
    \centering
    \caption{Input variables of ParT.}
    \label{tab:input_var}
  	\resizebox{1.\textwidth}{!}{
        \begin{tabular}{cl}
        
        \hline
        Object level & Observable variables \\
        
        \hline
        \multirow{3}{*}{\makecell{Reconstructed\\ particle}}
        
        & 4-momentum ($E, p_x, p_y, p_z$) \\
        & Direction ($\theta, \phi$) \\
        & Number of tracks and clusters \\
        
        \hline
        \multirow{4}{*}{Track}
        & Number of hits \\
        & Endpoint position \\
        & 3-momentum ($|\vec{p}|$, $p_x$, $p_y$, $p_z$, $p_T$) \\
        & ${\rm d}E/{\rm d}x$ (mean of 5--85\% truncation and quartiles) \\
        
        \hline
        \multirow{14}{*}{Cluster}
        & Number of hits \\
        & Energy \\
        & Position of shower starting point \\
        & Position of center of gravity \\
        & Fractal dimension~\cite{FD_PRL} \\
        & Second moment ($M_2$) \\
        & Distance between ECAL inner surface and shower starting point \\
        & Distance between ECAL inner surface and center of gravity \\
        & Distance between ECAL inner surface and the innermost hit \\
        & Distance between ECAL inner surface and the outermost hit \\
        & Maximum distance between cluster hits and the track helix (for charged particles) \\
        & Maximum distance between cluster hits to the axis from the innermost hit to the center of gravity \\
        & Average distance between cluster hits to the axis from the innermost hit to the center of gravity \\
        
        & Hit time spectrum (the fastest time and quintiles) \\
        
        \hline
        \multirow{4}{*}{\makecell{Closest\\ charged\\ cluster}}
        
        & Minimum distance between cluster hits of each other \\
        & Number of hits \\
        & Energy \\
        & Ratio of $E_{\rm cluster}$ to $p_{\rm track}$   \\
        
        \hline
        \end{tabular}
    }
\end{table*}

\subsubsection{Fake particle suppression}

\begin{figure}[htbp]
    \centering
    \includegraphics[width=.45\textwidth]{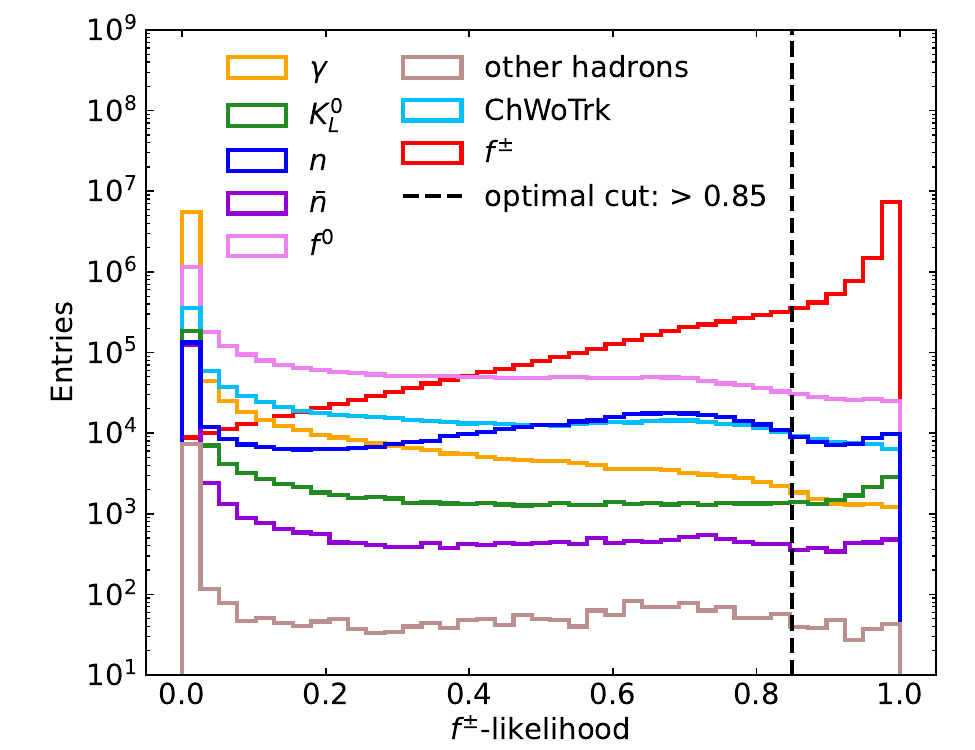}
    \caption{Likelihoods of fake particles from charged shower fragments ($f^{\pm}$) of different types of reconstructed neutral particles. $f^{0}$ represents fake particles from neutral shower fragments. The dashed line corresponds to the optimal working point to minimize the BMR by excluding the identified $f^{\pm}$.}
    \label{fig:likelihood}
\end{figure}

Fake particles are those extra reconstructed neutral particles arising from shower fragments of charged particles, denoted by $f^{\pm}$.
Fig.~\ref{fig:likelihood} compares the output $f^{\pm}$-likelihoods of different types of reconstructed neutral particles.
Identified $f^{\pm}$ is defined as those reconstructed neutral ones with $f^{\pm}$-likelihood larger than some threshold.
By excluding the identified $f^{\pm}$, the optimal BMR of 2.75\% is derived when $f^{\pm}$-likelihood $>$ 0.85.

\bibliographystyle{elsarticle-num} 
\bibliography{ref}

\begin{thebibliography}{10}
\expandafter\ifx\csname url\endcsname\relax
  \def\url#1{\texttt{#1}}\fi
\expandafter\ifx\csname urlprefix\endcsname\relax\def\urlprefix{URL }\fi
\expandafter\ifx\csname href\endcsname\relax
  \def\href#1#2{#2} \def\path#1{#1}\fi

\bibitem{Evans:2008zzb}
{LHC Machine}, JINST 3 (2008) S08001.
\newblock \href {https://doi.org/10.1088/1748-0221/3/08/S08001}
  {\path{doi:10.1088/1748-0221/3/08/S08001}}.

\bibitem{BelleII_Phy_2019}
W.~Altmannshofer, et~al., {The Belle II Physics Book}, PTEP 2019~(12) (2019)
  123C01, [Erratum: PTEP 2020, 029201 (2020)].
\newblock \href {http://arxiv.org/abs/1808.10567} {\path{arXiv:1808.10567}},
  \href {https://doi.org/10.1093/ptep/ptz106} {\path{doi:10.1093/ptep/ptz106}}.

\bibitem{BelleII_TDR}
T.~Abe, et~al., {Belle II Technical Design Report} (11 2010).
\newblock \href {http://arxiv.org/abs/1011.0352} {\path{arXiv:1011.0352}}.

\bibitem{BESIII_Phy}
C.-Z. Yuan, S.~L. Olsen, {The BESIII physics programme}, Nature Rev. Phys.
  1~(8) (2019) 480--494.
\newblock \href {http://arxiv.org/abs/2001.01164} {\path{arXiv:2001.01164}},
  \href {https://doi.org/10.1038/s42254-019-0082-y}
  {\path{doi:10.1038/s42254-019-0082-y}}.

\bibitem{BESIII:2009fln}
M.~Ablikim, et~al., {Design and Construction of the BESIII Detector}, Nucl.
  Instrum. Meth. A 614 (2010) 345--399.
\newblock \href {http://arxiv.org/abs/0911.4960} {\path{arXiv:0911.4960}},
  \href {https://doi.org/10.1016/j.nima.2009.12.050}
  {\path{doi:10.1016/j.nima.2009.12.050}}.

\bibitem{ILC_TDR_Sum}
{The International Linear Collider Technical Design Report - Volume 1:
  Executive Summary} (6 2013).
\newblock \href {http://arxiv.org/abs/1306.6327} {\path{arXiv:1306.6327}}.

\bibitem{CLIC_CDR}
{Physics and Detectors at CLIC: CLIC Conceptual Design Report} (2 2012).
\newblock \href {http://arxiv.org/abs/1202.5940} {\path{arXiv:1202.5940}},
  \href {https://doi.org/10.5170/CERN-2012-003}
  {\path{doi:10.5170/CERN-2012-003}}.

\bibitem{CEPC_TDR_Acc}
W.~Abdallah, et~al., {CEPC Technical Design Report: Accelerator}, Radiat.
  Detect. Technol. Methods 8~(1) (2024) 1--1105.
\newblock \href {http://arxiv.org/abs/2312.14363} {\path{arXiv:2312.14363}},
  \href {https://doi.org/10.1007/s41605-024-00463-y}
  {\path{doi:10.1007/s41605-024-00463-y}}.

\bibitem{CEPC_CDR_Phy}
M.~Dong, et~al., {CEPC Conceptual Design Report: Volume 2 - Physics \&
  Detector} (11 2018).
\newblock \href {http://arxiv.org/abs/1811.10545} {\path{arXiv:1811.10545}}.

\bibitem{FCC:2018evy}
A.~Abada, et~al., {FCC-ee: The Lepton Collider}: {Future Circular Collider
  Conceptual Design Report Volume 2}, Eur. Phys. J. ST 228~(2) (2019) 261--623.
\newblock \href {https://doi.org/10.1140/epjst/e2019-900045-4}
  {\path{doi:10.1140/epjst/e2019-900045-4}}.

\bibitem{European_Strategy_2020}
E.~S. Group, Deliberation document on the 2020 update of the european strategy
  for particle physics., \url{https://cds.cern.ch/record/2735179} (2020).

\bibitem{US_P5}
U.~P5, Report of the particle physics project prioritization panel (p5).,
  \url{https://www.usparticlephysics.org/wp-content/uploads/2018/03/FINAL\_P5\_Report\_053014.pdf}
  (2014).

\bibitem{Pandora}
J.~S. Marshall, M.~A. Thomson, {Pandora Particle Flow Algorithm}, in:
  {International Conference on Calorimetry for the High Energy Frontier}, 2013,
  pp. 305--315.
\newblock \href {http://arxiv.org/abs/1308.4537} {\path{arXiv:1308.4537}}.

\bibitem{Arbor}
M.~Ruan, H.~Videau, {Arbor, a new approach of the Particle Flow Algorithm}, in:
  {International Conference on Calorimetry for the High Energy Frontier}, 2013,
  pp. 316--324.
\newblock \href {http://arxiv.org/abs/1403.4784} {\path{arXiv:1403.4784}}.

\bibitem{CMS_PFA}
A.~M. Sirunyan, et~al., {Particle-flow reconstruction and global event
  description with the CMS detector}, JINST 12~(10) (2017) P10003.
\newblock \href {http://arxiv.org/abs/1706.04965} {\path{arXiv:1706.04965}},
  \href {https://doi.org/10.1088/1748-0221/12/10/P10003}
  {\path{doi:10.1088/1748-0221/12/10/P10003}}.

\bibitem{ALEPH_PFA}
D.~Buskulic, et~al., {Performance of the ALEPH detector at LEP}, Nucl. Instrum.
  Meth. A 360 (1995) 481--506.
\newblock \href {https://doi.org/10.1016/0168-9002(95)00138-7}
  {\path{doi:10.1016/0168-9002(95)00138-7}}.

\bibitem{Brient:2004yq}
J.~C. Brient, {Improving the jet reconstruction with the particle flow method:
  An introduction}, in: {11th International Conference on Calorimetry in
  High-Energy Physics (Calor 2004)}, 2004, pp. 445--451.

\bibitem{Videau:2002sk}
H.~Videau, J.~C. Brient, {Calorimetry optimised for jets}, in: {10th
  International Conference on Calorimetry in High Energy Physics (CALOR 2002)},
  2002, pp. 747--760.

\bibitem{ILC_TDR_Det}
H.~Abramowicz, et~al., {The International Linear Collider Technical Design
  Report - Volume 4: Detectors} (6 2013).
\newblock \href {http://arxiv.org/abs/1306.6329} {\path{arXiv:1306.6329}}.

\bibitem{LHC_upgrade_CMS}
{Technical Proposal for the Phase-II Upgrade of the CMS Detector} (6 2015).
\newblock \href {https://doi.org/10.17181/CERN.VU8I.D59J}
  {\path{doi:10.17181/CERN.VU8I.D59J}}.

\bibitem{Kakati:2024dun}
N.~Kakati, E.~Dreyer, A.~Ivina, F.~A. Di~Bello, L.~Heinrich, M.~Kado, E.~Gross,
  {HGPflow: Extending Hypergraph Particle Flow to Collider Event
  Reconstruction}{\href{https://arxiv.org/abs/2410.23236}{https://arxiv.org/abs/2410.23236}}
  (10 2024).
\newblock \href {http://arxiv.org/abs/2410.23236} {\path{arXiv:2410.23236}}.

\bibitem{Pata:2023rhh}
J.~Pata, E.~Wulff, F.~Mokhtar, D.~Southwick, M.~Zhang, M.~Girone, J.~Duarte,
  {Improved particle-flow event reconstruction with scalable neural networks
  for current and future particle detectors}, Commun. Phys. 7~(1) (2024) 124.
\newblock \href {http://arxiv.org/abs/2309.06782} {\path{arXiv:2309.06782}},
  \href {https://doi.org/10.1038/s42005-024-01599-5}
  {\path{doi:10.1038/s42005-024-01599-5}}.

\bibitem{DiBello:2022iwf}
F.~A. Di~Bello, et~al., {Reconstructing particles in jets using set transformer
  and hypergraph prediction networks}, Eur. Phys. J. C 83~(7) (2023) 596.
\newblock \href {http://arxiv.org/abs/2212.01328} {\path{arXiv:2212.01328}},
  \href {https://doi.org/10.1140/epjc/s10052-023-11677-7}
  {\path{doi:10.1140/epjc/s10052-023-11677-7}}.

\bibitem{Thomson:2009rp}
M.~A. Thomson, {Particle Flow Calorimetry and the PandoraPFA Algorithm}, Nucl.
  Instrum. Meth. A 611 (2009) 25--40.
\newblock \href {http://arxiv.org/abs/0907.3577} {\path{arXiv:0907.3577}},
  \href {https://doi.org/10.1016/j.nima.2009.09.009}
  {\path{doi:10.1016/j.nima.2009.09.009}}.

\bibitem{Ruan:2018yrh}
M.~Ruan, et~al., {Reconstruction of physics objects at the Circular Electron
  Positron Collider with Arbor}, Eur. Phys. J. C 78~(5) (2018) 426.
\newblock \href {http://arxiv.org/abs/1806.04879} {\path{arXiv:1806.04879}},
  \href {https://doi.org/10.1140/epjc/s10052-018-5876-z}
  {\path{doi:10.1140/epjc/s10052-018-5876-z}}.

\bibitem{Yu:2020bxh}
D.~Yu, M.~Ruan, V.~Boudry, H.~Videau, J.-C. Brient, Z.~Wu, Q.~Ouyang, Y.~Xu,
  X.~Chen, {The measurement of the $H\rightarrow \tau \tau $ signal strength in
  the future $e^{+}e^{-}$ Higgs factories}, Eur. Phys. J. C 80~(1) (2020) 7.
\newblock \href {https://doi.org/10.1140/epjc/s10052-019-7557-y}
  {\path{doi:10.1140/epjc/s10052-019-7557-y}}.

\bibitem{Zhao:2018jiq}
H.~Zhao, Y.-F. Zhu, C.-D. Fu, D.~Yu, M.-Q. Ruan, {The Higgs Signatures at the
  CEPC CDR Baseline}, Chin. Phys. C 43~(2) (2019) 023001.
\newblock \href {http://arxiv.org/abs/1806.04992} {\path{arXiv:1806.04992}},
  \href {https://doi.org/10.1088/1674-1137/43/2/023001}
  {\path{doi:10.1088/1674-1137/43/2/023001}}.

\bibitem{Yu:2017mpx}
D.~Yu, M.~Ruan, V.~Boudry, H.~Videau, {Lepton identification at particle flow
  oriented detector for the future $e^{+}e^{-}$ Higgs factories}, Eur. Phys. J.
  C 77~(9) (2017) 591.
\newblock \href {http://arxiv.org/abs/1701.07542} {\path{arXiv:1701.07542}},
  \href {https://doi.org/10.1140/epjc/s10052-017-5146-5}
  {\path{doi:10.1140/epjc/s10052-017-5146-5}}.

\bibitem{Yu:2021pxc}
D.~Yu, T.~Zheng, M.~Ruan, {Lepton identification performance in Jets at a
  future electron positron Higgs Z factory} (5 2021).
\newblock \href {http://arxiv.org/abs/2105.01246} {\path{arXiv:2105.01246}},
  \href {https://doi.org/10.1088/1748-0221/16/06/P06013}
  {\path{doi:10.1088/1748-0221/16/06/P06013}}.

\bibitem{An:2018jtk}
F.~An, S.~Prell, C.~Chen, J.~Cochran, X.~Lou, M.~Ruan, {Monte Carlo study of
  particle identification at the CEPC using TPC dE / dx information}, Eur.
  Phys. J. C 78~(6) (2018) 464.
\newblock \href {http://arxiv.org/abs/1803.05134} {\path{arXiv:1803.05134}},
  \href {https://doi.org/10.1140/epjc/s10052-018-5803-3}
  {\path{doi:10.1140/epjc/s10052-018-5803-3}}.

\bibitem{Zhu:2022hyy}
Y.~Zhu, S.~Chen, H.~Cui, M.~Ruan, {Requirement analysis for dE/dx measurement
  and PID performance at the CEPC baseline detector}, Nucl. Instrum. Meth. A
  1047 (2023) 167835.
\newblock \href {http://arxiv.org/abs/2209.14486} {\path{arXiv:2209.14486}},
  \href {https://doi.org/10.1016/j.nima.2022.167835}
  {\path{doi:10.1016/j.nima.2022.167835}}.

\bibitem{ParT}
H.~Qu, C.~Li, S.~Qian, {Particle Transformer for Jet Tagging} (2 2022).
\newblock \href {http://arxiv.org/abs/2202.03772} {\path{arXiv:2202.03772}}.

\bibitem{Qu:2019gqs}
H.~Qu, L.~Gouskos, {Jet Tagging via Particle Clouds}, Phys. Rev. D 101~(5)
  (2020) 056019.
\newblock \href {http://arxiv.org/abs/1902.08570} {\path{arXiv:1902.08570}},
  \href {https://doi.org/10.1103/PhysRevD.101.056019}
  {\path{doi:10.1103/PhysRevD.101.056019}}.

\bibitem{vaswani2017attention}
A.~Vaswani, {Attention is all you need}, Advances in Neural Information
  Processing Systems (2017).

\bibitem{CMS-DP-2024-066}
\href{https://cds.cern.ch/record/2904702}{{A unified approach for jet tagging
  in Run 3 at $\sqrt{s}$=13.6 TeV in CMS}}~(CMS-DP-2024-066) (2024).
\newline\urlprefix\url{https://cds.cern.ch/record/2904702}

\bibitem{CMS:2024okz}
A.~Hayrapetyan, et~al., {Measurement of inclusive and differential cross
  sections of single top quark production in association with a W boson in
  proton-proton collisions at $\sqrt{s}$ = 13.6 TeV} (9 2024).
\newblock \href {http://arxiv.org/abs/2409.06444} {\path{arXiv:2409.06444}}.

\bibitem{Brigljevic:2024vuv}
V.~Brigljevic, et~al., {HHH Whitepaper}, in: {HHH Workshop}, 2024.
\newblock \href {http://arxiv.org/abs/2407.03015} {\path{arXiv:2407.03015}}.

\bibitem{JOI}
H.~Liang, Y.~Zhu, Y.~Wang, Y.~Che, C.~Zhou, H.~Qu, M.~Ruan, {Jet-Origin
  Identification and Its Application at an Electron-Positron Higgs Factory},
  Phys. Rev. Lett. 132~(22) (2024) 221802.
\newblock \href {http://arxiv.org/abs/2310.03440} {\path{arXiv:2310.03440}},
  \href {https://doi.org/10.1103/PhysRevLett.132.221802}
  {\path{doi:10.1103/PhysRevLett.132.221802}}.

\bibitem{GSHCAL_NIMA}
P.~Hu, Y.~Wang, D.~Du, Z.~Hua, S.~Qian, C.~Fu, Y.~Liu, M.~Ruan, J.~Wang,
  Y.~Wang, {GSHCAL at future e+e\ensuremath{-} Higgs factories}, Nucl. Instrum.
  Meth. A 1059 (2024) 168944.
\newblock \href {https://doi.org/10.1016/j.nima.2023.168944}
  {\path{doi:10.1016/j.nima.2023.168944}}.

\bibitem{Manqi_CEPCWS_Chicago_2019}
M.~Ruan, Simulation and performance of the cepc: at cdr and recent update,
  \url{https://indico.cern.ch/event/820586/contributions/3553194/attachments/1909878/3155585/1-Simulation-and-Performance-Chicago.pdf}
  (2019).

\bibitem{Lai:2021rko}
P.-Z. Lai, M.~Ruan, C.-M. Kuo, {Jet performance at the circular
  electron-positron collider}, JINST 16~(07) (2021) P07037.
\newblock \href {http://arxiv.org/abs/2104.05029} {\path{arXiv:2104.05029}},
  \href {https://doi.org/10.1088/1748-0221/16/07/P07037}
  {\path{doi:10.1088/1748-0221/16/07/P07037}}.

\bibitem{Hu:2023dbm}
P.~Hu, Y.~Wang, D.~Du, Z.~Hua, S.~Qian, C.~Fu, Y.~Liu, M.~Ruan, J.~Wang,
  Y.~Wang, {GSHCAL at future e+e\ensuremath{-} Higgs factories}, Nucl. Instrum.
  Meth. A 1059 (2024) 168944.
\newblock \href {https://doi.org/10.1016/j.nima.2023.168944}
  {\path{doi:10.1016/j.nima.2023.168944}}.

\bibitem{Liang:2024hox}
H.~Liang, L.~Li, Y.~Zhu, X.~Shen, M.~Ruan, {Measurement of CKM element
  $|V_{cb}|$ from $W$ boson decays at the future Higgs factories} (6 2024).
\newblock \href {http://arxiv.org/abs/2406.01675} {\path{arXiv:2406.01675}}.

\bibitem{Li:2022tlo}
X.~Li, M.~Ruan, M.~Zhao, {Prospect for measurement of the CP-violating
  phase~$\phi _s$ in the $B_{s}\rightarrow J/\psi \phi $ channel at a future Z
  factory}, Eur. Phys. J. C 84~(8) (2024) 859.
\newblock \href {http://arxiv.org/abs/2205.10565} {\path{arXiv:2205.10565}},
  \href {https://doi.org/10.1140/epjc/s10052-024-13217-3}
  {\path{doi:10.1140/epjc/s10052-024-13217-3}}.

\bibitem{Zhao:2022lyl}
Z.~Zhao, S.~Yang, M.~Ruan, M.~Liu, L.~Han, {Measurement of the effective weak
  mixing angle at the CEPC*}, Chin. Phys. C 47~(12) (2023) 123002.
\newblock \href {http://arxiv.org/abs/2204.09921} {\path{arXiv:2204.09921}},
  \href {https://doi.org/10.1088/1674-1137/acf91f}
  {\path{doi:10.1088/1674-1137/acf91f}}.

\bibitem{Zheng:2020qyh}
T.~Zheng, J.~Wang, Y.~Shen, Y.-K.~E. Cheung, M.~Ruan, {Reconstructing $K^0_S$
  and $\Lambda $ in the CEPC baseline detector}, Eur. Phys. J. Plus 135~(3)
  (2020) 274.
\newblock \href {https://doi.org/10.1140/epjp/s13360-020-00272-4}
  {\path{doi:10.1140/epjp/s13360-020-00272-4}}.

\bibitem{CEPC_Snowmass}
H.~Cheng, et~al., {The Physics potential of the CEPC. Prepared for the US
  Snowmass Community Planning Exercise (Snowmass 2021)}, in: {Snowmass 2021},
  2022.
\newblock \href {http://arxiv.org/abs/2205.08553} {\path{arXiv:2205.08553}}.

\bibitem{Manqi_HKIAS_2024}
M.~Ruan, Advanced reconstruction: Pfa \& jet origin identification.,
  \url{https://indico.cern.ch/event/1335278/contributions/5733469/attachments/2785402/4856212/Advanced%20Reco%20-%20PFA%20-%20Jet%20origin%20-%20POST.pdf}
  (2024).

\bibitem{Manqi_CSI_talk}
M.~Ruan, Advanced reconstruction of hadronic events using ai.,
  \url{https://indico.cern.ch/event/1439509/timetable/#26-advanced-reconstruction-of}
  (2025).

\bibitem{GSHCAL_talk}
H.~Li, Progress of the cepc gs-hcal.,
  \url{https://indico.ihep.ac.cn/event/22941/contributions/171374/attachments/84638/108004/CLHCP2024\_CEPC\_GSHCAL.pdf}
  (2024).

\bibitem{Liang:2022qxn}
H.~Liang, Y.~Zhu, P.-Z. Lai, M.~Ruan, {Optimization of tracker configuration
  for the CEPC}, JINST 17~(11) (2022) P11016.
\newblock \href {http://arxiv.org/abs/2209.00397} {\path{arXiv:2209.00397}},
  \href {https://doi.org/10.1088/1748-0221/17/11/P11016}
  {\path{doi:10.1088/1748-0221/17/11/P11016}}.

\bibitem{Whizard}
W.~Kilian, T.~Ohl, J.~Reuter, {WHIZARD: Simulating Multi-Particle Processes at
  LHC and ILC}, Eur. Phys. J. C 71 (2011) 1742.
\newblock \href {http://arxiv.org/abs/0708.4233} {\path{arXiv:0708.4233}},
  \href {https://doi.org/10.1140/epjc/s10052-011-1742-y}
  {\path{doi:10.1140/epjc/s10052-011-1742-y}}.

\bibitem{Pythia6}
T.~Sjostrand, S.~Mrenna, P.~Z. Skands, {PYTHIA 6.4 Physics and Manual}, JHEP 05
  (2006) 026.
\newblock \href {http://arxiv.org/abs/hep-ph/0603175}
  {\path{arXiv:hep-ph/0603175}}, \href
  {https://doi.org/10.1088/1126-6708/2006/05/026}
  {\path{doi:10.1088/1126-6708/2006/05/026}}.

\bibitem{MoradeFreitas:2004sq}
P.~Mora~de Freitas, \href{https://core.ac.uk/download/pdf/46779218.pdf}{{Mokka,
  main guidelines and future}}, in: {International Conference on Linear
  Colliders (LCWS 04)}, 2004, pp. 441--444.
\newline\urlprefix\url{https://core.ac.uk/download/pdf/46779218.pdf}

\bibitem{Mokka_CEPCNote}
C.~Fu,
  \href{http://cepcdoc.ihep.ac.cn/DocDB/0001/000167/001/cepc-sim.pdf}{{Full
  Simulation Software at CEPC}}, CEPC Document Server (2017).
\newline\urlprefix\url{http://cepcdoc.ihep.ac.cn/DocDB/0001/000167/001/cepc-sim.pdf}

\bibitem{Bierlich:2022pfr}
C.~Bierlich, et~al., {A comprehensive guide to the physics and usage of PYTHIA
  8.3}, SciPost Phys. Codeb. 2022 (2022) 8.
\newblock \href {http://arxiv.org/abs/2203.11601} {\path{arXiv:2203.11601}},
  \href {https://doi.org/10.21468/SciPostPhysCodeb.8}
  {\path{doi:10.21468/SciPostPhysCodeb.8}}.

\bibitem{Alwall:2014hca}
J.~Alwall, R.~Frederix, S.~Frixione, V.~Hirschi, F.~Maltoni, O.~Mattelaer,
  H.~S. Shao, T.~Stelzer, P.~Torrielli, M.~Zaro, {The automated computation of
  tree-level and next-to-leading order differential cross sections, and their
  matching to parton shower simulations}, JHEP 07 (2014) 079.
\newblock \href {http://arxiv.org/abs/1405.0301} {\path{arXiv:1405.0301}},
  \href {https://doi.org/10.1007/JHEP07(2014)079}
  {\path{doi:10.1007/JHEP07(2014)079}}.

\bibitem{Bellm:2015jjp}
J.~Bellm, et~al., {Herwig 7.0/Herwig++ 3.0 release note}, Eur. Phys. J. C
  76~(4) (2016) 196.
\newblock \href {http://arxiv.org/abs/1512.01178} {\path{arXiv:1512.01178}},
  \href {https://doi.org/10.1140/epjc/s10052-016-4018-8}
  {\path{doi:10.1140/epjc/s10052-016-4018-8}}.

\bibitem{FD_PRL}
M.~Ruan, D.~Jeans, V.~Boudry, J.-C. Brient, H.~Videau, {Fractal Dimension of
  Particle Showers Measured in a Highly Granular Calorimeter}, Phys. Rev. Lett.
  112~(1) (2014) 012001.
\newblock \href {http://arxiv.org/abs/1312.7662} {\path{arXiv:1312.7662}},
  \href {https://doi.org/10.1103/PhysRevLett.112.012001}
  {\path{doi:10.1103/PhysRevLett.112.012001}}.

\end{thebibliography}






\end{document}